\documentclass[prx,aps,superscriptaddress,twocolumn,footinbibe,longbibliography]{revtex4-1}
\pdfoutput=1
\usepackage[colorlinks=true,urlcolor=blue]{hyperref}
\usepackage{amsfonts}
\usepackage{graphicx}
\usepackage{placeins}
\usepackage{amsmath}
\usepackage{amssymb}
\usepackage{tabularx}
\usepackage{xcolor}
\usepackage{color}
\usepackage{bbold}
\usepackage{bm}
\usepackage[normalem]{ulem}
\usepackage{braket}
\usepackage{enumitem}

\definecolor{red}{rgb}{0.75,0,0}
\definecolor{blue}{rgb}{0,0,0.75}
\definecolor{green}{rgb}{0,0.5,0}

\DeclareMathOperator{\sign}{sign}

\DeclareRobustCommand{\DIEP}{\ensuremath{%
    \mathchoice{\includegraphics[height=2ex]{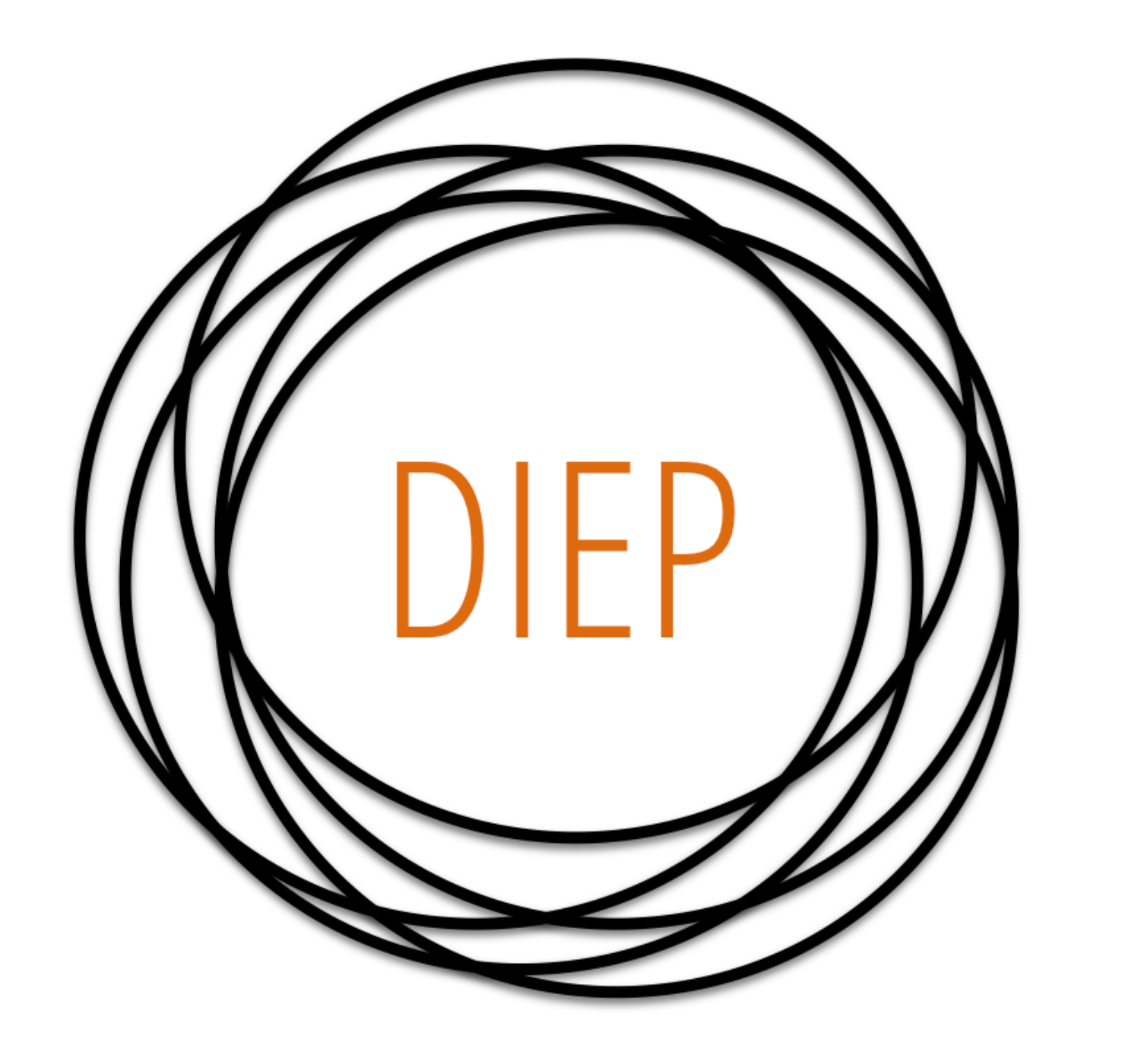}}
    {\includegraphics[height=2ex]{DIEPs.pdf}}
    {\includegraphics[height=1.5ex]{DIEPs.pdf}}
    {\includegraphics[height=1ex]{DIEPs.pdf}}
  }}

\begin{document}

\title{Topological waves in passive and active fluids on curved surfaces: a unified picture}

\author{Richard Green}
\affiliation{Institute for Theoretical Physics, University of Amsterdam, 1090 GL Amsterdam, The Netherlands}
\affiliation{Instituut-Lorentz, Universiteit Leiden, P.O. Box 9506, 2300 RA Leiden, The Netherlands}
\author{Jay Armas}
\affiliation{Institute for Theoretical Physics, University of Amsterdam, 1090 GL Amsterdam, The Netherlands}
\affiliation{\DIEP Dutch Institute for Emergent Phenomena (DIEP), University of Amsterdam, 1090 GL Amsterdam, The Netherlands}
\author{Jan de Boer}
\affiliation{Institute for Theoretical Physics, University of Amsterdam, 1090 GL Amsterdam, The Netherlands}
\affiliation{\DIEP Dutch Institute for Emergent Phenomena (DIEP), University of Amsterdam, 1090 GL Amsterdam, The Netherlands}
\author{Luca Giomi}
\affiliation{Instituut-Lorentz, Universiteit Leiden, P.O. Box 9506, 2300 RA Leiden, The Netherlands}

\begin{abstract}
We investigate the occurrence of topologically protected waves in classical fluids confined on curved surfaces. Using a combination of topological band theory and real space analysis, we demonstrate the existence of a system-independent mechanism behind topological protection in two-dimensional passive and active fluids. This allows us to formulate an index theorem linking the number of modes, determined by the topology of Fourier space, to the real space topology of the surface on which they are hosted. With this framework in hand, we review two examples of topological waves in two-dimensional fluids, namely oceanic shallow-water waves propagating on the Earth's rotating surface and momentum waves in active polar fluids spontaneously ``flocking'' on substrates endowed with a ${\rm U}(1)$ isometry (e.g. surfaces of revolution). Our work suggests some simple rules to engineer topological modes on surfaces in passive and active soft matter systems.
\end{abstract}

\maketitle

\section{Introduction}

The discovery of topologically protected modes in condensed matter systems, has unquestionably represented one of the major breakthroughs in theoretical physics of the past decades. Topology explains the origin of the quantization of the Hall conductance, where time-reversal symmetry is broken by a magnetic field \cite{vonKlitzing1980,Thouless1982,Haldane1988}, as well as the existence of unidirectional edge modes that are insensitive to various type of disorder in topological insulators \cite{Kane2005,Bernevig2006,Fu2007,Hasan2010,Moore2010,Ryu2010}. Aside from the quantum world, topological mechanisms have been reported in a plethora of classical mechanical systems, such as networks of springs \cite{Kane2013}, hinges \cite{Chen2014,Paulose2015}, pendula \cite{Huber2016}, gyroscopes \cite{Nash2015} etc.

Yet, it is in the realm of classical fluids that the notion of topological protection offers some of its most spectacular manifestations. As recently demonstrated by Delplace {\em et al.} \cite{Delplace2017,Perrot2019}, certain types of oceanic waves, known in Earth science as Kelvin and Yanai waves and responsible for large scale geophysical phenomena such as the El Ni\~no Southern Oscillation, are topologically trapped along the equator owing to the combination of the Earth's rotation, which breaks time-reversal symmetry, and closed topology. Additionally, analogous topologically protected modes arise in active fluids, such as ``flocks'' of self-propelled agents collectively moving on curved substrates \cite{Shankar2017}. In this case, the fluctuations in the flock density and flying direction conspire with the substrate spatial curvature, resulting in a spectral structure featuring topologically distinct bands depending on the geometry of the surface, characterized by different integer-valued Chern numbers $C_{n}=1/(2\pi)\int_{\rm BZ} d^{2}k\,\Omega_{n}(\bm{k})$, where $\Omega_{n}$ is the so called Berry curvature associated with the $n-$th band and the integral is computed over the first Brillouin zone.  

These findings open new avenues in the vibrant field of topological matter by: {\em 1)} highlighting the role of classical fluids as an ideal testing ground to identify new realizations of topological mechanisms; {\em 2)} suggesting the possibility of an interplay between Fourier space and real space topology; {\em 3)} shedding light on the role of curvature as a control parameter to engineer specific topological modes, with potential applications to microfluidic technologies.

In this article, we aim at giving a unified description of topological modes in two-dimensional passive and active classical fluids constrained to lie on a curved substrate $\mathcal{M}$. After a short review of topological waves (Sec. \ref{sec: approach}), we demonstrate that the main physical mechanisms behind the existence of topological waves in two-dimensional fluids can be recovered, in system-independent manner, from the so called shallow water equations (Sec. \ref{sec: shallow water general}) on surfaces endowed with a ${\rm U(1)}$ isometry in a specific direction $\phi$ (e.g. surfaces of revolution). The existence of this isometry allows for equilibrium flows along $\phi$, which naturally break time-reversal symmetry. This results in a gauge-invariant component of the substrate spin connection $A_{\phi}$ that serves as the waves' effective mass, thus influencing the waves' band structure. This mechanism depends solely on the emergence of directed motion in the background flow and is insensitive to the specific mechanism fuelling the flow, thus rendering the occurrence of topological modes in oceanic waves and flocks, two different manifestations of the same fundamental physical mechanism. This generality reflects on the band structure at the edge (i.e. where the mass changes sign) in real space, where the net number of modes (i.e. the difference in the number of upstream and downstream traveling modes), is given by:
\begin{equation}\label{eq:index_theorem_1}
\mathcal{N} = \chi(\mathcal{M})\;,
\end{equation}
in which $\chi$ is the Euler characteristic of the surface $\mathcal{M}$. This, in turn, is related to a set of Chern numbers in the bulk through the universal relation:
\begin{equation}\label{eq:index_theorem_2}
\sum_{n \in \mathbb{0}(A_{\phi})} \Delta C_{n} = \chi(\mathcal{M})\;, 	
\end{equation}
where the sum runs over the set $\mathbb{0}(A_{\phi})$ of all the isolated zeros of $A_{\phi}$, $\Delta C_{n}$ is the difference between the Chern numbers of the lower band across each isolated zero as a function of transverse coordinates, and in Eq. \eqref{eq:index_theorem_2} we used the bulk-edge correspondence \cite{Kane20133}.

In Secs. \ref{sec: shallow water} and \ref{sec: TT(u)}, we provide a direct proof that the differential equations governing the propagation of shallow water waves on a rotating sphere and momentum density perturbations in a flock can be cast in the same general form and we also review the bulk-edge correspondence in Sec. \ref{sec: bulk-edge}. Conclusions will be finally drawn in Sec. \ref{sec: conclusions}.

\section{Topological modes}
\label{sec: approach}

\begin{figure*}[t]
\centering
    \includegraphics[width=1.0\textwidth]{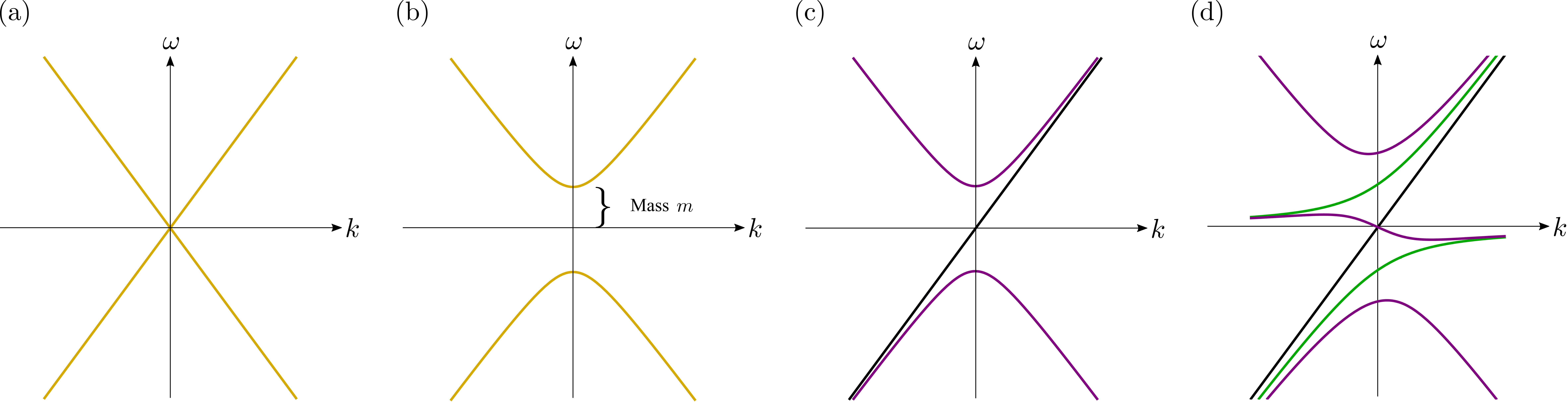}
\caption{Dispersion relations for: (a) bulk Hamiltonian with time reversal symmetry; (b) bulk Hamiltonian with band gap; (c) edge Hamiltonian for Dirac fermion with $m(y) \sim y$; and (d) edge Hamiltonian for shallow water waves on rotating sphere, also with $m(y) \sim y$ where the Kelvin wave is shown in black and the Yanai wave in green. Bands for negative frequency $\omega$ against wavenumber $k$ in each case can be obtained by $(\omega,k) \to (-\omega,- k)$.}
 \label{fig: fig1}
\end{figure*}

In this section, we review some fundamental concepts about topological modes in quantum and classical systems and outline two complementary approaches to unveil their mathematical and physical properties that will be used in this article: the Chern number analysis in the bulk and the direct solution for bounded modes on the edge.

Let us consider a system whose departure from a steady state configuration can be described in term of a state vector $\bm{\Psi}=\bm{\Psi}(\bm{r},t)$. In a quantum system, $\bm{\Psi}$ is a Hilbert space vector, whereas in classical fluids, $\bm{\Psi}$ expresses the difference between an arbitrary set of hydrodynamic variables with respect to their steady state configuration (e.g. $\bm{\Psi}=[\delta\rho,\delta\bm{v},\ldots]$, with $\rho$ and $\bm{v}$ the local density and velocity of the fluid). Let us further assume $\bm{\Psi}$ to obey a partial differential equation of the from: $\partial_{t}\bm{\Psi}=\bm{F}(\bm{\Psi},\nabla\bm{\Psi})$, with $\bm{F}$ a function that vanishes for $\bm{\Psi}=\bm{0}$. Sufficiently close to the steady state, thus for small $|\bm{\Psi}|$ values, the dynamics can be described by the linear equation:
\begin{equation}\label{eq:schroedinger}
\partial_{t}\Psi_{i} = \mathcal{H}_{ij}\Psi_{j}\;,
\end{equation}	
where $\mathcal{H}_{ij}=\partial F_{i}/\partial\Psi_{j}|_{\bm{\Psi}=\bm{0}}$ plays the role of an effective Hamiltonian dictating evolution of the perturbation $\bm{\Psi}$ in time. Looking for plane waves solutions of Eq. \eqref{eq:schroedinger} of the form $\bm{\Psi}=\bm{\hat{\Psi}}\exp i(\omega t-\bm{k}\cdot\bm{r})$ yields: 
\begin{equation}\label{eq:omega}
i\omega \hat{\Psi}_{i} = \hat{\mathcal{H}}_{ij}\hat{\Psi}_{j}\;,
\end{equation}
where $\bm{\hat{\Psi}}=\bm{\hat{\Psi}}(\bm{\bm{k},\omega})$ is the amplitude of the Fourier mode of wave vector $\bm{k}$ and frequency $\omega$ and $\bm{\mathcal{\hat{H}}}=\bm{\mathcal{\hat{H}}}(\bm{k})$ the Fourier-transformed Hamiltonian. Now, if the dynamics is completely reversible, Eq. \eqref{eq:schroedinger} has the typical structure of a conservation law and the spectrum of the perturbation $\bm{\Psi}$ is characterized by a linear dispersion relation, $\omega\sim \pm k$, corresponding to two bands crossing each other at $\bm{k}=\bm{0}$, as shown in Fig. \ref{fig: fig1}a. In the presence of irreversible processes (i.e. processes that are not invariant under time-reversal symmetry), however, the spectrum may partition into topologically distinct bands. To illustrate this concept, let us consider the following Hamiltonian:
\begin{equation}\label{eq:dirac_hamiltonian}
\bm{\mathcal{\hat{H}}}
= 
\begin{bmatrix}
im & -k_{x}+i k_{y} \\
k_{x} + i k_{y} & -im	
\end{bmatrix}\;,
\end{equation}
where $m$ is an arbitrary constant parameter. Such a Hamiltonian could be derived, for instance, for the $(2+1)-$dimensional Dirac equation, describing the dynamics of a relativistic spin$-1/2$ fermion \cite{Jackiw1976}, in which case $m$ represents the fermion mass, i.e.:
\begin{equation}\label{eq:dirac}
(i\gamma^{\mu}\partial_{\mu}-m)\bm{\Psi} = \bm{0}\;,	
\end{equation}
where $\mu=t,\,x,\,y$ and $\gamma^{\mu}$ are proportional to the Pauli matrices: $\gamma^{0}=-\sigma_{z}$, $\gamma^{1}=-i\sigma_{x}$ and $\gamma^{2}=-i\sigma_{y}$, with $\sigma_{i}\sigma_{j}=i\epsilon_{ijk}\sigma_{k}$, with $\epsilon_{ijk}$ the antisymmetric symbol. From Eq. \eqref{eq:dirac_hamiltonian} one can readily compute the dispersion relation:
\begin{equation}\label{eq:gap}
\omega = \pm \sqrt{k^{2}+m^{2}}\;,
\end{equation}
where $k^{2}=k_{x}^{2}+k_{y}^{2}$. Evidently, Eq. \eqref{eq:gap}, implies the existence of a band gap at $\bm{k}=\bm{0}$, unless $m=0$, see Fig. \ref{fig: fig1}b. 

We are in fact interested in situations with a boundary or internal interface, where the system architecture or physical properties change and so consider a position-dependent mass, $m=m(\bm{r})$, that remains approximately constant away from these interfaces, but varies sharply near $m=0$. If the bands have different Chern numbers across the interfaces where the band gap at $\bm{k}=\bm{0}$ closes, these peripheral or internal ``edges'' become the locations of uni-directional trapped waves, whose propagation in the bulk is forbidden by the topological fingerprint of the band structure.

To demonstrate this latter statement we consider the case in which $m=m(y)$ depends only on the coordinate $y$ and analyze the topological invariants associated with the band structure entailed in the Hamiltonian Eq. \eqref{eq:dirac_hamiltonian}, valid away from $y=0$ where $m$ is approximately constant. For each eigenvector of the Hamiltonian we calculate the Berry curvature:
\begin{equation}
\Omega_{n}(\bm{k}) = \nabla_{\bm{k}} \times \bm{A}_{n}\;, 
\end{equation}
where $\nabla_{\bm{k}}=[\partial_{k_{x}},\partial_{k_{y}}]$ and $\bm{A}_{n}$ is the so called Berry connection associated with the $n-$th eigenvector:
\begin{equation}\label{eq:berry_connection}
\bm{A}_{n} = i\bm{\Psi}_{n}^{*}\cdot(\nabla_{\bm{k}}\bm{\Psi}_{n})\;.	
\end{equation}
The integral of $\Omega_{n}$ over the first Brillouin zone yields an integer valued Chern number, whose value changes at $y=0$, where $m$ changes in sign. By virtue of the 
bulk-edge correspondence, relating the net number and direction of modes localized at the ``edge" to the change in Chern number in the bulk, we can then infer the direction and net number of possible bounded edge modes at $y=0$ without further calculations \cite{PhysRevLett.71.3697, Graf_2013}. Further details are provided in Sec. \ref{sec: bulk-edge}.

In order for the Chern number to be well defined, it is necessary to regularize the short distance behavior. This can be done by adding second order differential terms to the Dirac Hamiltonian \cite{Bal_2019} leading to an additional term $\epsilon k^{2}$, with $\epsilon$ a cut-off. With this modification, the Hamiltonian Eq. \eqref{eq:dirac_hamiltonian} becomes:
\begin{equation}\label{eq:regulated_hamiltonian}
\bm{\mathcal{\hat{H}}}
= 
\begin{bmatrix} 
i\left(m - \epsilon k^2\right) & -k_x + i k_y \\ 
k_x + i k_y & -i\left(m -\epsilon k^2\right) 
\end{bmatrix}\;,
\end{equation}
whose eigenvalues readily yield the dispersion relation:
\begin{equation}
\omega = \pm \sqrt{(m-\epsilon k^2)^2 + k^2}\;.
\end{equation}
Next, expressing the wave vector $\bm{k}$ in polar coordinates $(k,\phi)$ and introducing the isotropic function $\theta=\theta(k) \in [0,\pi]$, such that:
\begin{equation}
\cos \theta = \frac{m-\epsilon k^2}{\sqrt{\left(m-\epsilon k^{2}\right)^2 + k^2}}\;,
\end{equation}
we can recast Eq. \eqref{eq:regulated_hamiltonian}, up to an overall factor, in the form:
\begin{equation}
\bm{\mathcal{\hat{H}}}
= 
\begin{bmatrix}  
i \cos{\theta} & - \sin{\theta}\,e^{-i \phi} \\ 
\sin{\theta}\,e^{i \phi} & -i \cos{\theta} 
\end{bmatrix}\;,
\end{equation}
whose eigenvectors are given by:
\begin{equation}\label{eq:dirac_eigenvectors}
\bm{\Psi}_+ = 
\begin{bmatrix} 
\cos{\frac{1}{2}\theta} \\ 
-i e^{i \phi} \sin{\frac{1}{2}\theta} 
\end{bmatrix}\;,
\qquad 
\bm{\Psi}_- = 
\begin{bmatrix} 
\sin{\frac{1}{2}\theta} \\ 
i e^{i \phi} \cos{\frac{1}{2}\theta} \
\end{bmatrix}\;.
\end{equation}
Notice that, for $k=0$, $\cos\theta=\sign(m)$, whereas in the limit $k\to\infty$, $\cos\theta \to \sign(-\epsilon)$.
Now, using Eqs. \eqref{eq:berry_connection} and \eqref{eq:dirac_eigenvectors} yields:
\begin{equation}
\bm{A}_{\pm} 
= \frac{\pm\cos\theta-1}{2 k}\,\bm{\hat{\phi}}\;,
\end{equation}
from which the corresponding Chern number is readily calculated as:
\begin{align}
\label{eq: Chern number}
C_{\pm} 
&= \frac{1}{2\pi} \int_{0}^{2\pi}{\rm d}\phi \int_{0}^{\infty}{\rm d}k\,\partial_{k}\left(\pm\frac{1}{2}\cos\theta \right) \notag \\[5pt]
&= \mp \frac{\sign(m) + \sign(\epsilon)}{2}\;.
\end{align}
Now, since $\sign(\cdots)= \pm 1$, as long the sign of $m$ is constant, $C_{\pm}$ is also an integer constant. Yet, at $y=0$, where $m$ changes in sign, $C_{\pm}$ changes by $\pm 1$. According to the bulk-edge correspondence, the number of modes propagating along such an interface toward the left ($n_{L}$) and the right $(n_{R})$ is related to the Chern number difference across the interface (since $C_+ = - C_-$, it is sufficient to consider only $C_-$):
\begin{equation}
\Delta C_{-} = n_R - n_L\;.
\end{equation}
Suppose that $m'(0)>0$, so that $m(y)>0$ for $y>0$ and $m(y)<0$ for $y<0$. Crossing the interface from $y>0$ to $y<0$, implies $\Delta C_- =-1$, so $n_R - n_L = -1$ i.e. there exist one extra mode travelling left, which in this case corresponds to the $x>0$ direction. When $m'(0)<0$ on the other hand, $\Delta C_- = 1$, and there is one extra mode travelling right i.e. in the direction $x<0$. Summarizing, depending on the details of $m(y)$, $\Delta C_- = \pm1$ and this result is independent of the regulator $\epsilon$.

The argument above relied on $m=$ constant with opposite sign on either side of an interface and we saw that $C_\pm$ depended only on the sign (not magnitude) of $m$. In the case of $m$ varying with $y$, one can think of extending the argument to a small patch at any point in the bulk where the patch is small enough to consider $m$ to be constant. In that case, a low $k$ cut-off (in addition to the high $k$ $\epsilon$ cut-off above) should be taken into account corresponding to the patch under consideration. This can be implemented by introducing a lower limit $k_{\rm low}$ in the integral in Eq. \eqref{eq: Chern number}. However as the interface $m=0$ is approached in real space, $k_{\rm low} \to \infty$ and this allows $C_\pm$ to interpolate between the two values it takes deep in the bulk either side of the boundary.

With the Dirac fermion, as in the classical fluid systems we will study next, the previous topological construction can be supported by an edge analysis near $y=0$ where we seek a direct solution of the wave equations for the bounded modes as a function of $y$. Thus, we look for solutions of Eq. \eqref{eq:dirac} of the form:
\begin{equation}
\bm{\Psi}(x,y,t) = 
\begin{bmatrix}
\psi_{x}(y) \\
\psi_{y}(y)	
\end{bmatrix}	
e^{i(\omega t-k_{x}x)}\;.
\end{equation}
Then, taking $\psi_{\pm}=\psi_{x}\pm i\psi_{y}$ in Eq. \eqref{eq:dirac} yields:
\begin{equation}\label{eq:dirac_matrix}
\partial_{y}
\begin{bmatrix}	
\psi_{+}\\
\psi_{-}	
\end{bmatrix}
=
\begin{bmatrix}
- m & k+\omega \\
k-\omega & m	
\end{bmatrix}
\begin{bmatrix}	
\psi_{+}\\
\psi_{-}	
\end{bmatrix}~.
\end{equation}
Now, close to $y=0$, we can approximate $m \approx m'(0)y$, where $(\cdots)'=\partial_{y}(\cdots)$, and look for {\em bounded} solutions, such that:
\begin{equation}
\lim_{y\rightarrow\pm\infty} \psi_{\pm} = 0\;.	
\end{equation}
For $\omega=\pm k$, this yields:
\begin{equation}
\begin{bmatrix}
\psi_{\pm} \\
\psi_{\mp} 
\end{bmatrix}
= 
\psi_{0}
\begin{bmatrix}
e^{-\frac{1}{2}\,|m'(0)| y^2}\\
0
\end{bmatrix}\;,
\end{equation}
with $\psi_{0}$ a constant. The direction of propagation of the bounded mode is determined by the sign of $m'(0)$: when $m'(0)>0$, then $\psi_+$ is bounded provided that $\psi_-=0$ and $\omega= k$. For $m'(0)<0$ on the other hand, $\psi_-$ is now the bounded mode provided that $\psi_+=0$ and $\omega = -k$. Thus:
\begin{equation}
\bm{\Psi}(x,y,t) 
= \frac{\psi_{0}}{2}
\begin{bmatrix}
1 \\
\pm i
\end{bmatrix}
e^{i\omega (t \pm x)-\frac{1}{2}\,|m'(0)| y^2}\;.
\end{equation}
This corresponds to a topologically protected mode propagating unidirectionally along the positive (negative) $x-$direction for $\omega= k$ ($\omega = -k$) where the direction is determined by $\sign[m'(0)]$.

There are also higher harmonics for $\omega^2 \ne k^2$. Eliminating $\psi_-$ from Eq. \eqref{eq:dirac_matrix} gives the second-order equation:
\begin{equation}
\frac{\psi''_+}{\psi_+} = \left[ m'(0) y \right]^{2} - \left[\omega^2 - k^2 + m'(0) \right]\;,
\end{equation}
whose bounded solutions are given by:
\begin{equation} \label{eq: psi+}
\psi_+ = H_{n+(s+1)/2}(\sqrt{|m'(0)|} y)\,e^{i(\omega t - k x)-\frac{1}{2}\,|m'(0)| y^2}\;
\end{equation}
where $s=\sign[m'(0)]$ and $H_n (...)$ are the Hermite polynomials, provided that the dispersion relation
\begin{equation}
\omega^2 = k^2 + (2 n + 2) |m'(0)|\;, \qquad n = 0,\,1,\,2\ldots
\end{equation}
\noindent is satisfied. These higher harmonics, shown for $n=0$ along with the $\omega = k$ mode in Fig. \ref{fig: fig1}c in the case $m'(0)>0$, may propagate in both directions $x\lessgtr 0$. Thus for any value of $\omega$, the total number of modes travelling in the $x>0$ direction ($\partial \omega / \partial k >0$) is always one more than the number of modes travelling in the $x<0$ direction ($\partial \omega / \partial k <0$). In the case $m'(0)<0$, there is instead one extra mode travelling in the $x<0$ direction. We note that the individual number of modes $n_R$ and $n_L$ is dependent on the details of $m(y)$. For instance, Ref.~\cite{Tauber_2019} considered the case of a sharp interface and found the modes with frequency $\omega=\sign[m'(0)] k$ but not the higher harmonics; this can also be seen in Eq. \eqref{eq: psi+} since $\psi_+ \to 0$ as $|m'(0)| \to \infty$.  Regardless, the net number of edge modes is $\mathcal{N}=\sign[m'(0)]$ in agreement with the topological analysis.

\section{Topological waves on surfaces with U(1) isometry}
\label{sec: shallow water general}

\subsection{Transport equation}

Let us consider a two-dimensional fluid of density $\rho$ and velocity $\bm{v}$, constrained to lie on curved surface $\mathcal{M}$ characterized by a ${\rm U(1)}$ isometry, namely a surface that is invariant with respect to translation, rotation or dilatation along a specific direction. Let us further assume the surface to be parametrized in terms of a pair of coordinates, i.e. $\bm{r}=\bm{r}(\theta,\phi)$, such that $\bm{g}_{i}=\partial_{i}\bm{r}$, with $i=\theta,\,\phi$, is a basis vector on the tangent plane and $g_{ij}=\bm{g}_{i}\cdot\bm{g}_{j}$ the associated metric tensor. For simplicity, here we assume the coordinates are everywhere orthogonal, so that the surface can be parametrized in the simple form $\bm{r}/L=[x,y,z]$, with $L$ a generic length scale setting the system size and
\begin{equation}\label{eq:surface}
\left\{
\begin{array}{l}
x = \xi(\theta) \cos\phi \\[5pt]
y = \xi(\theta) \sin\phi \\[5pt]
z = \eta(\theta)
\end{array}
\right.\;,
\end{equation}
where $\phi$ is the coordinate in the direction of the isometry and $\xi$ and $\eta$ dimensionless functions of $\theta$. The surface line element is then:
\begin{equation}\label{eq:metric surface}
{\rm d}s^{2} = L^{2}(p\,{\rm d}\theta^{2}+ r\,{\rm d}\phi^{2})\;,
\end{equation}
where $p=\xi'^{2}+\eta'^{2}$, $r=\xi^{2}$ and the prime indicates differentiation with respect to $\theta$. In Appendix \ref{sec: App shallow water}, we will consider the general case of a surface endowed with a ${\rm U(1)}$ isometry, but lacking of globally orthogonal coordinates. 

\begin{figure}[t]
\centering
\includegraphics[width=0.4\columnwidth]{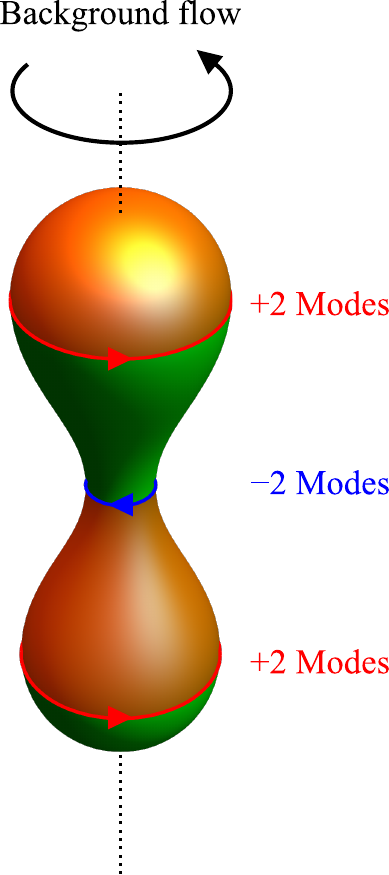}
\caption{Location of topological modes of shallow water waves traveling on background flow rotating counter-clockwise on a surface of revolution. Orange and green shades indicate regions of the surface having opposite signed spin connection $A_{\phi}$, being the waves' effective mass $m$.}
 \label{fig: general shape}
\end{figure}

Let us further assume the dynamics of the material fields $\rho$ and $\bm{v}=v^{i}\bm{g}_{i}$ to be governed by the following set of partial differential equations:
\begin{subequations}\label{eq:generic_shallow_water}
\begin{gather}
\partial_{t}\rho + \nabla\cdot(\rho\bm{v}) = 0\;,\\
\partial_{t}\bm{v} + \bm{v}\cdot\nabla\bm{v} + g\nabla\rho = 0\;~,
\end{gather}
\end{subequations}
where $g$ is a coupling constant and all differential operators are covariant (e.g. $\nabla\cdot\bm{v}=g_{ij}\nabla^{j}v^{i}$). Despite being very simple, Eqs. \eqref{eq:generic_shallow_water} describe, at the linear order in the velocity gradients, wave propagation in a vast range of physical systems. In addition to the quiescent state $\rho={\rm const}$ and $\bm{v}=\bm{0}$, Eqs. \eqref{eq:generic_shallow_water} admit a non-trivial time-independent solution endowed with the same ${\rm U(1)}$ symmetry of the substrate. The latter corresponds to a flow rotating at constant angular velocity $\Omega$ along the $\phi-$direction and whose density varies across the surface, namely:
\[
\rho_{0} = \langle \rho \rangle \left(1+\frac{r-\left\langle r \right\rangle}{2\alpha} \right)\;,\qquad
v_0^\theta = 0\;,\qquad
v_0^\phi = \Omega\;,
\]
where $\langle \rho \rangle$ is the average density, $\alpha = g \langle\rho\rangle/(L^2 \Omega^2)$ and $\left\langle r \right\rangle$ is the average of the function $r$ across the surface. Details of this and subsequent calculations are provided in Appendix \ref{sec: App shallow water}. Now, taking $\delta\rho=\rho-\rho_{0}$ as well as $\delta v^{i}=v^{i}-v^{i}_{0}$ and introducing the following dimensionless variables:
\begin{gather*}
X = 2\sqrt{\frac{r}{\alpha}}\,\phi\,,\quad 
Y = 2\sqrt{\frac{p}{\alpha}}\,\theta\,,\quad
T = 2\Omega t\,, \\
U = \frac{1}{\Omega}\sqrt{\frac{r}{\alpha}}\,\delta v^{\phi}\;,\quad
V = \frac{1}{\Omega}\sqrt{\frac{p}{\alpha}}\,\delta v^{\theta}\;,\quad
H = \frac{\delta\rho}{\rho_{0}}\;,
\end{gather*}
and linearizing Eqs. \eqref{eq:generic_shallow_water} about the steady state solution (i.e. $U=V=H=0$), yields
\begin{equation}
\label{eq: X Y general metric}
\partial_{T} 
\begin{bmatrix} 
H\\ 
U\\
V 
\end{bmatrix} 
=
\bm{\mathcal{H}}
\begin{bmatrix}
H \\ 
U \\ 
V
\end{bmatrix}\;,
\end{equation}
where we have introduced the effective Hamiltonian:
\begin{equation}\label{eq:h_u1}
\bm{\mathcal{H}}
=
\begin{bmatrix} 
0 & - \partial_X & - \partial_Y - B \left(r+\Delta\right) \\
-\partial_X & 0 & m \\ 
-\partial_Y & -m & 0
\end{bmatrix}\;,
\end{equation}
with
\begin{gather*}
m = -\frac{r'}{2\Delta}\;,\qquad
\Delta = \sqrt{pr}\;,\\
B = \frac{\sqrt{\alpha}}{2\Delta\sqrt{p+r+2\Delta}} \left(\frac{r'}{2\alpha}+ \frac{\Delta'}{\Delta} \right)\;.
\end{gather*}
In Eq. \eqref{eq:h_u1} the quantity $m$ is a mass-like parameter that plays an analogous role to the mass $m$ in Eq. \eqref{eq:dirac_hamiltonian} and is responsible for the gap in the spectrum of the perturbation. Furthermore, as we detail in Appendix \ref{sec: App shallow water}:
\begin{equation}\label{eq:spin_connection}
m = A_{\phi}\;,	
\end{equation}
where $A_{\phi}$ is the component of the spin-connection $A_{i}=\bm{e}_{\theta}\cdot\partial_{i}\bm{e}_{\phi}$, with $\bm{e}_{i}=\bm{g}_{i}/|\bm{g}_{i}|$ and $i=\theta,\,\phi$, in the direction of the isometry \footnote{The precise spatially covariant statement is $m=k^i A_i$ where $k^i\partial_i=\partial_\phi$ is the Killing vector field associated with the $U(1)$ isometry. Equilibrium requires that $\mathcal{L}_k A_i=0$ where $\mathcal{L}_k$ is the Lie derivative along $k^i$. This condition is trivially satisfied. However, the spin-connection transforms as a $U(1)$ gauge field under rotations of the tangent vectors such that $A_i\to A_i+\partial_i \bar{\Lambda}$ for some arbitrary function $\bar{\Lambda}$. Under this transformation, equilibrium requires that $k^i\partial_i \bar{\Lambda}=0$. Thus the scalar $k^iA_i$ is gauge-invariant and $m$ is fixed due to the presence of the isometry.}. Considering length-scales $L$ sufficiently large, the $B$ term in Eq. \eqref{eq:h_u1} is at least of order $\mathcal{O}\left(L^{-1}\right)$ and can be ignored compared to $\partial_{X}$ and $\partial_{Y}$ \footnote{Formally, we must introduce coordinates that render the metric flat $Y\to L\sqrt{\alpha} Y/2$ and $X\to L\sqrt{\alpha} X/2$ and consider $L$ to be very large such that $\Omega L$ is kept finite. Thus the $\beta$-plane approximation is more accurate for slowly rotating configurations.}. This is consistent with the understanding of hydrodynamics as an effective theory where curvature effects are subleading in a derivative expansion \cite{Bhattacharyya:2012nq, Banerjee:2017ouw}. In this limit, we can expect topological modes when the spin connection in the direction of the isometry changes sign. As illustrated in Fig. \ref{fig: general shape}, this occurs along a finite number of parallels of $\mathcal{M}$, where the surface normal vector $\bm{n}=(\bm{g}_{\theta}\times\bm{g}_{\phi})/|\bm{g}_{\theta}\times\bm{g}_{\phi}|$ is orthogonal to the axis of rotation (the $z-$axis in this case). These parallels are closed {\em geodesic} lines on the surface and, consistently with the generic argument given in Sec. \ref{sec: approach}, can be thought as internal interfaces that partition the substrate in regions characterized by a different intrinsic geometry.

To be more precise, we note that the function $m$ near any point $Y_0$ can be Taylor expanded such that $m=m(Y_0) +m'(Y_0)(Y-Y_0)+...$. Introducing flat coordinates \cite{Note2}, $m'(Y_0)\sim \mathcal{O}\left(L^{-1}\right)$, for sufficiently small scales the subleading term in the Taylor expansion can be ignored, except near the interface where $m(Y_0=0)=0$. Thus, taking into account that $B$ is also subleading, we have shown that away from the interfaces the Hamiltonian \eqref{eq:h_u1} reduces to the common $f-$plane approximation where $m$ is approximately constant, while near the interface we recover the $\beta-$plane approximation where $m$ is a linear function of $y$. In other words, in the hydrodynamic limit and for surfaces with a $U(1)$ isometry, we uncover universal physics near the interfaces. As the bulk Chern numbers associated with \eqref{eq:h_u1} have been calculated in \cite{Delplace2017, Souslov2017, Tauber_2019} in the $f-$plane approximation, we deduce that $\Delta C_n= \pm 2$ for the lower band across any of the $n$ zeros of $m=A_\phi$, with the sign or direction of the modes given by $\sign[m'(0)]$.

Compared to the Dirac fermion (see Sec. \ref{sec: approach}), the system described by Eqs. \eqref{eq:generic_shallow_water} has an extra topological bounded mode at $m=0$ arising from the additional degree of freedom. Systems with two degrees of freedom, such as the Dirac fermion, have typically two bands for each $k$ value, whereas systems with three degrees of freedom have three. Thus, crossing the interface where $m$ changes in sign implies $\Delta C = 2$, instead of $\Delta C = 1$ as in the case of the Dirac fermion.

\subsection{Index theorem for topological modes}

\begin{table*}[t]
\centering
\begin{ruledtabular}  
\renewcommand{\arraystretch}{2}
\begin{tabular}{cccc}
Sphere & Torus & Catenoid & Helicoidal sphere \\
\hline\\ 
\includegraphics[width=0.2\textwidth]{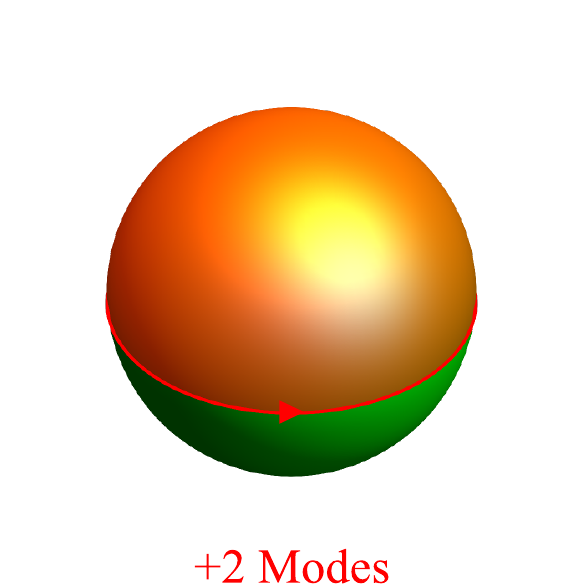} &
\includegraphics[width=0.2\textwidth]{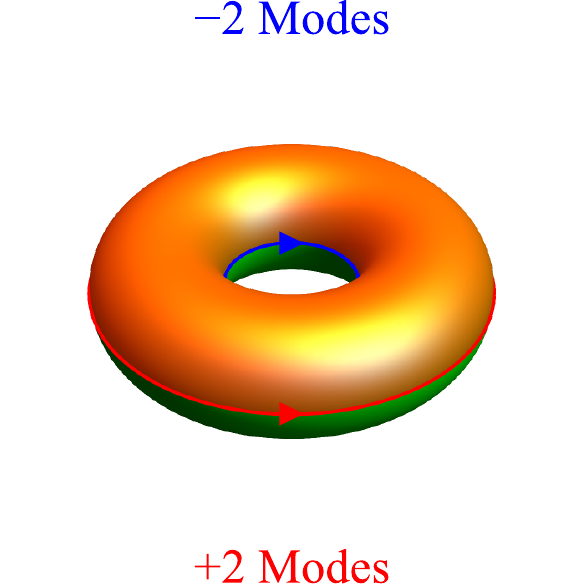} &
\includegraphics[width=0.2\textwidth]{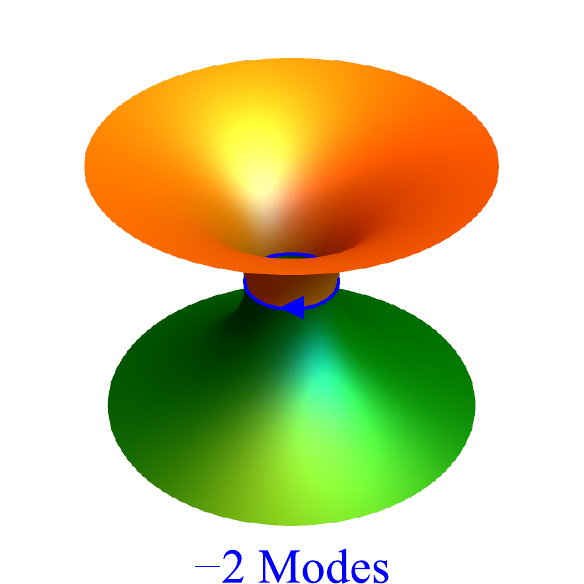} &
\includegraphics[width=0.2\textwidth]{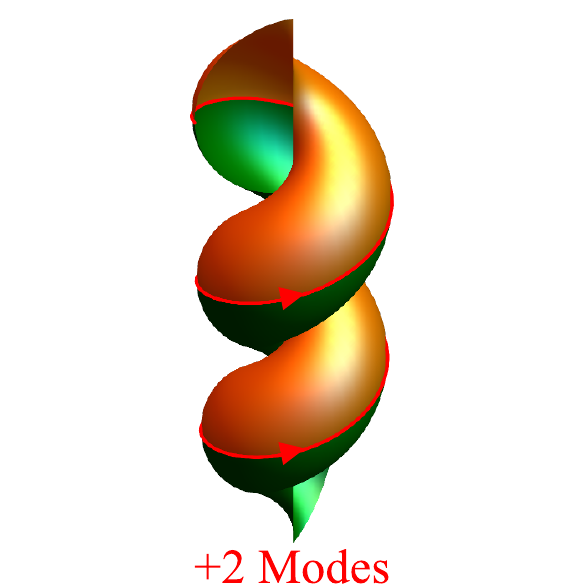} \\
\hline
$\left\{
\begin{array}{l}
x = \sin\theta\cos\phi\\[-5pt]
y = \sin\theta\sin\phi\\[-5pt]
z = \cos\theta
\end{array}
\right.$ &
$\left\{
\begin{array}{l}
x = (\varrho+\cos\theta)\cos\phi\\[-5pt]
y = (\varrho+\cos\theta)\sin\phi\\[-5pt]
z = \sin\theta
\end{array}
\right.$ &
$\left\{
\begin{array}{l}
x = \cosh(\theta/c)\cos\phi\\[-5pt]
y = \cosh(\theta/c)\sin\phi\\[-5pt]
z = \theta
\end{array}
\right.$ &
$\left\{
\begin{array}{l}
x = \sin\theta\cos\phi\\[-5pt]
y = \sin\theta\sin\phi\\[-5pt]
z = \cos\theta+\phi/\pi
\end{array}
\right.$ \\
\hline
$\theta \in [0,\pi)\quad \phi \in [0,2\pi)$ &
$\theta \in [0,2\pi)\quad \phi \in [0,2\pi)$ &
$\theta \in (-\infty,\infty)\quad \phi \in [0,2\pi)$ &
$\theta \in (-\infty,\infty)\quad \phi \in [0,2\pi)$ \\
\hline
$m=-\cos\theta$ & 
$m=\sin\theta$ & 
$m=-\tanh\theta$ & 
$m=-\pi\sin 2\theta [(\pi^{2}-1)\sin^{2}\theta+1]^{-1/2}$ \\
\end{tabular}
\end{ruledtabular}
\caption{\label{fig: shapes_modes}Summary of topological modes on various surfaces shown with a counterclockwise direction of rotation. On the sphere the mass vanishes at $\theta_{0}=\pi/2$ and $m'(\theta_{0})=1$, resulting in net two counterclockwise propagating modes along the equator. On the torus of aspect ratio $\varrho \ge 1$, the mass vanishes at $\theta_{0}=0$ and $\theta_{1}=\pi$ and  $m'(\theta_{0})=-m'(\theta_{1})=1$. Thus, there are net two counterclockwise propagating modes on the external equator and net two clockwise propagating modes on the internal equator. On the catenoid the mass vanishes at $\theta_{0}=0$ and $m'(\theta_{0})=-1$, resulting in net two clockwise propagating modes along the equator. On the helicoidal sphere (for which the vertical height $z=\phi/\pi + \cos \theta$), there are net two counterclockwise modes propagating along the helix $\theta_0 = \pi/2$, where $m(\theta_0)=0$ and $m'(\theta_0)=2$.}
\end{table*}

The identification of the parallel geodesics, where $m=A_{\phi}$ changes in sign, as internal interfaces where topological modes can propagate, allows us to demonstrate the main results of this paper, Eq. \eqref{eq:index_theorem_1} and \eqref{eq:index_theorem_2}, relating three classes of topological invariants in real and Fourier space, namely the Chern number of the bulk bands, the net number of bounded modes at the edges, and the Euler characteristic $\chi(\mathcal{M})$.

Geodesic parallels on surfaces of revolution are circles whose distance from the $z-$axis is either maximal or minimal compared to neighboring points (i.e. $r'=0$). Geometrically, the former correspond to circular ``ridges'', revolving around the surface, and the latter to ``valleys'' (Fig. \ref{fig: general shape}). If we take the direction of rotation $\Omega$ to be counterclockwise, since their direction of propagation is given by $\sign(m')$, topological modes travel clockwise on valleys and counterclockwise on ridges. Thus the {\em net} number $\mathcal{N}$ of topological modes propagating parallel to the direction of rotation can be expressed as the difference:
\begin{equation}
\mathcal{N} = 2\left( \#\,\text{ridges}-\#\,\text{valleys}\right)\;,
\end{equation}
where we have taken into account that each ridge and valley sustains two topological modes according to the analysis in the previous section. Now, in the framework of Morse theory (see e.g. Ref. \cite{Milnor1963}), ridges and valleys can be assigned an index $i$ associated with the structure of the function $F=F(X,Y)$ representing their distance from an arbitrary plane passing through the surface's axis of revolution, hence $F=\xi$ for the class of surfaces described by Eq. \eqref{eq:surface}. Thus valleys have index $i_{\rm valley}=1$, whereas $i_{\rm ridge}=2$~\footnote{In general, a Morse function $F:\mathcal{M}_{d}\rightarrow\mathbb{R}$ on a $d-$dimensional manifold $\mathcal{M}_{d}$, is a real-valued smooth function whose critical points are not degenerate. In the neighbourhood of the critical point $\bm{r}_{0}$, the function can then be expanded at the quadratic order as: $F(\bm{r})=F(\bm{r}_{0})-x_{1}^{2}\cdots- x_{i}^{2} + x_{i+1}^{2}\cdots + x_{d}^{2}$, with $(x_{1},\,x_{2},\,\ldots\,x_{d})$ local coordinates. The integer $i$ is then defined as the index of $F$ at the critical point $\bm{r}_{0}$. For a two-dimensional manifold: $F(\bm{r})=F(\bm{r}_{0}) \pm x_{1}^{2}\pm x_{2}^{2}$, where $(+,+)$ denote a minimum of $F$, $(-,-)$ a maximum and $(+,-)$ and $(-,+)$ a saddle point. Thus minima have index $i=0$, saddle points have $i=1$ and maxima have $i=2$.}. Moreover, Morse's theorem requires that:
\begin{equation}\label{eq:morse_theorem}
\sum_{n}(-1)^{i_{n}}J_{i_{n}} = \chi(M)\;,
\end{equation} 
where $J_{i_{n}}$ is twice the number of geodesic parallels having index $i_{n}$. Since parallel geodesics do not exist outside of ridges and valleys, $J_{1}=2\#\,\text{valleys}$ and $J_{2}=2\#\,\text{ridges}$, the total net number of modes $\mathcal{N}$ satisfies
\begin{equation}
\mathcal{N} = \chi(\mathcal{M}).
\end{equation}
\noindent Finally, we may rewrite \eqref{eq:morse_theorem} in terms of change in Chern numbers across each isolated zero of $A_\phi$ (as a function of the transverse coordinate) such that 
\begin{equation}
\sum_{n \in \mathbb{0}(A_{\phi})} \Delta C_{n} = \chi(\mathcal{M})\;,
\end{equation}
where $\Delta C_{n}$ denotes the change in the Chern number of the lower band $C_-$ across each zero of $A_\phi$. In summary, the net number of topological modes, taking into account their direction of propagation, propagating according to Eqs. \eqref{eq:generic_shallow_water} on a closed surface of revolution, is itself topological and equal to the Euler characteristic of the surface.

In Table \ref{fig: shapes_modes} we illustrate how the above result applies to the sphere and the axisymmetric torus. We also show the catenoid and the helicoidal sphere which are not compact and so do not fulfil the conditions behind Eq. \eqref{eq:index_theorem_1}. In the case of a geodesically complete (i.e. without boundaries) catenoid, $\mathcal N = 1/(2\pi) \int {\rm d}A\,K=-2$, with $K$ the Gaussian curvature. Finally, the helicoidal sphere is neither compact nor smooth as it is singular at $\theta=0, \pi$, but has net $\mathcal{N}=2$ modes localised on the helix given by $\theta=\pi/2$.

\section{Application to oceanic shallow water waves}
\label{sec: shallow water}

In this section, we show how the general framework introduced in Sec. \ref{sec: shallow water general} can be applied to oceanic shallow water waves, in such a way to recover the seminal result by Delplace {\em et al}. \cite{Delplace2017}. In particular, since in Sec.~\ref{sec: shallow water general} we uncovered universal physics, the edge analysis that we perform in this context applies to any of the isolated zeros of $A_\phi$ on an arbitrary surface of revolution.

The equations governing the dynamics of the water column, whose height $h \ll L$ is much smaller than the system size $L$, can be obtained from the incompressible Euler equation:
\begin{equation}\label{eq:eulers}
\rho(\partial_{t}+\bm{v}\cdot\nabla)\bm{v} = -\nabla p + \bm{f}\;,\qquad \nabla\cdot\bm{v} = 0\;,
\end{equation}
where $\rho$ and $\bm{v}=[v_{x},v_{y},v_{z}]$ are, respectively, the water density and velocity, $p$ is the pressure and $\bm{f}$ a body force. The large separation between the water depth and width, allows one to reduce the dimension of the problem by assuming the water column to be under hydrostatic balance, thus $p=\rho g h$, with $g$ the gravitational acceleration. Then, taking $v_{z}=0$ and integrating the horizontal components of the velocity field across $z$, i.e. $\bm{u}=\int_{0}^{h}{\rm d}z\,\bm{v}\approx \int_{0}^{\langle h \rangle}{\rm d}z\,\bm{v}$, with $\langle h \rangle$ the average depth, yields:
\begin{subequations}\label{eq:earth_shallow_water}
\begin{gather}
\partial_{t}h + \nabla\cdot(h\bm{u}) = 0\;,\\
(\partial_{t}+\bm{u}\cdot\nabla)\bm{u} = -g\nabla h + \rho^{-1}\bm{f}\;.   
\end{gather}
\end{subequations}
Earth's rotation breaks time-reversal symmetry by inducing the Coriolis force $\bm{f}=-2\rho\,\bm{\Omega}\times\bm{u}$, with $\bm{\Omega}$ the angular velocity of the Earth. Taking $\bm{\Omega}=\Omega\bm{\hat{z}}$ and linearizing Eqs. \eqref{eq:earth_shallow_water} about the rest state $\bm{u}=\bm{0}$ and $h=\langle h \rangle$ finally yields the following system of linear partial differential equations \cite{Matsuno1966,Vallis2017,Delplace2017}:
\begin{equation}\label{eqn: first}
\partial_{t}
\begin{bmatrix}
h \\
u_{x} \\
u_{y} \\
\end{bmatrix}
= \begin{bmatrix} 
0 & -\langle h \rangle \partial_x & -\langle h \rangle \partial_y \\ 
-g \partial_x & 0 & f \\ 
-g \partial_y & -f & 0 
\end{bmatrix} 
\begin{bmatrix} 
h \\ 
u_{x} \\ 
u_{y} 
\end{bmatrix},
\end{equation}
where $x$ is the coordinate in the surface in the direction of rotation (East), $y$ the coordinate in the surface orthogonal to $x$ (North) and $f=2 \Omega \sin y/R\approx \beta y$, with $R$ the Earth's radius and $\beta=2\Omega/R$, the Coriolis parameter. Eqs. \eqref{eqn: first} has evidently the same structure of Eqs. \eqref{eq:generic_shallow_water} with $m=f$ and $\rho=h$, thus, consistently with the general results given in Sec. \ref{sec: shallow water general} the system admits two eastward traveling topological modes localized along the equator (where $y=0$), corresponding to Kelvin and Yanai waves.

The structure of these equatorial waves can be further elucidated by a direct solution of Eqs. \eqref{eqn: first}. Following the strategy outlined in Sec. \ref{sec: approach} for the Dirac fermion, we first express Eqs. \eqref{eqn: first} in the following dimensionless units
\begin{gather*}
X = \frac{2\Omega}{\sqrt{g\langle h \rangle}}\,x\;,\quad
Y = \frac{2\Omega}{\sqrt{g\langle h \rangle}}\,y\;,\quad
T = 2\Omega\,t\;.\\
U = \frac{u_{x}}{\sqrt{g\langle h \rangle}}\;,\quad
V = \frac{u_{y}}{\sqrt{g\langle h \rangle}}\;,\quad
H = \frac{h}{\langle h \rangle}\;,\quad
F = \beta Y\;.
\end{gather*}
and look for plane wave solutions of the form $\exp i(\omega T-kX)$. This yields:
\begin{equation}
 \label{eqn: third}
\omega 
\begin{bmatrix} 
H \\ 
U \\ 
V \end{bmatrix} 
= 
\begin{bmatrix} 
0 & k & i \partial_Y \\ 
k & 0 & -i F \\ i \partial_Y & i F & 0 
\end{bmatrix} 
\begin{bmatrix} 
H \\ 
U \\ 
V 
\end{bmatrix},
\end{equation}
where the variables $(H,U,V)$ are functions of $Y$. Eliminating $U$ and rearranging gives
\begin{equation}\label{eqn: V eta}
\partial_Y 
\begin{bmatrix} 
V\\ 
H
\end{bmatrix} 
=  
\begin{bmatrix} 
\frac{kF}{\omega} & -i \omega - \frac{k^2}{i\omega} \\ 
-i \omega - \frac{F^2}{i\omega} & - \frac{k F}{\omega} 
\end{bmatrix} 
\begin{bmatrix} 
V \\ 
H  
\end{bmatrix}.
\end{equation}
Eq. \eqref{eqn: V eta} can now be solved with boundary conditions $\lim_{Y\rightarrow\pm\infty}(V,H)=0$ to find modes that are bounded at infinity. Rescaling $(X,Y,T) \to \beta^{1/2} (X,Y,T)$ and $H \to \beta^{-1/2} H$ and considering the case $\omega^2=k^2$ yields:
\begin{subequations}
\begin{gather}
V' = \pm Y V\;, \\
H' = \mp \frac{i(k^2-Y^2)}{k}\,V \mp Y H\;,
\end{gather}
\end{subequations}
where the $\pm$ correspond to $\omega=\pm k$ and $(\cdots)'=\partial_{Y}(\cdots)$. For $\omega = k$, there is a bounded mode for which $V=0$ and $ H \sim \exp(-Y^2/2)$, but there is no bounded solution for $\omega= - k$. This solution, known as Kelvin wave, represents therefore a dispersionless wave traveling only eastwards, with positive phase and group velocity $\partial \omega / \partial k >0$. 

By contrast, for $\omega^2 \ne k^2$, we can eliminate $H$ from Eq. \eqref{eqn: V eta} and obtain:
\begin{equation}
\frac{V''}{V} = -\omega^2 + k^2 + \frac{k}{\omega} + Y^2\;,  
\label{eq: modes}
\end{equation}
whose bounded solution are given by:
\begin{equation}
V = e^{-\frac{Y^2}{2}} H_n(Y) e^{i(\omega T - k X)}\;,
\end{equation}
where $H_n(Y)$ are the Hermite polynomials and $n$ is an integer solution of the equation:
\begin{equation}
\omega^2 - k^2 - \frac{k}{\omega} = 2 n +1\;.
\label{eq: disp rel}
\end{equation}
Equivalent expressions for $H$ and $U$ can be calculated from Eqs. \eqref{eqn: V eta} and \eqref{eqn: third} respectively. For $n=0$, there are two solutions satisfying the dispersion relation Eq. \eqref{eq: disp rel} for which also $\omega^2 \ne k^2$, namely:
\begin{equation}
\omega = \frac {k \pm \sqrt{k^2+4}}{2}\;.
\end{equation}
These are known as Yanai waves and, analogously to Kelvin waves, propagate eastward with positive group velocity. Finally, for $n\ge 1$, Eqs. \eqref{eq: disp rel} have three solutions consisting of two gravity waves and one lower frequency planetary or Rossby wave, which all can propagate both East and West. Fig. \ref{fig: fig1}d shows all the bands described for $n=0,\,1$. Since the dispersion relation Eq. \eqref{eq: disp rel} has the symmetry $(\omega,k) \to (-\omega,-k)$, the band structure for negative $\omega$ is obtained by reflecting in $\omega=0$ and $k=0$. This provides a complementary derivation of the results of Ref. \cite{196625}.

In summary, we find that the net number of edge modes for any given frequency is always $\mathcal{N}=2$, in agreement with the general analysis of Sec.~\ref{sec: shallow water general}. Even though we carried out this edge analysis near the equator of a sphere, it equally applies near any point for which $m=0$ in an arbitrary surface of revolution. We note that Ref.~\cite{Tauber_2019} performed a similar edge analysis in the case of a sharp interface and only found Kelvin- and Yanai-like waves. Again, regardless of the details on the interface, the net number of trapped modes is $\mathcal{N}=2$.

\section{Topological flocks on surfaces of revolution}
\label{sec: TT(u)}

In this section we demonstrate that the formalism developed in Sec. \ref{sec: shallow water general} and edge analysis of Sec. \ref{sec: shallow water} can also account for the existence of topological modes in ``flocks'' of self-propelled particles flowing on a surface of revolution \cite{Shankar2017}. The hydrodynamics of flocks in polar active fluids is traditionally described in terms of the phenomenological Toner-Tu equations \cite{Toner1995,Toner2005}, whose simplified form is given by:
\begin{subequations}\label{eq:toner_tu}
\begin{gather}
\partial_t \rho + \nabla\cdot(\rho \bm{v}) = 0\;, \\
(\partial_t  + \lambda \bm{v}\cdot\nabla)\bm{v} = \left(\alpha - \beta |\bm{v}|^{2} \right) \bm{v} - \frac{c^2}{\rho_{\rm eq}} \nabla \rho\;,
\end{gather}
\end{subequations}
with $\lambda$, $\alpha$, $\beta$ and $c$ phenomenological constants and $\rho_{\rm eq}$ is the density at a specific location on the surface. The field $\bm{v}$, in this context, is both a physical velocity and a broken symmetry variable and, for $\alpha>0$, the system described by Eqs. \eqref{eq:toner_tu} is characterized by a flocking steady state with $|\bm{v}|=v_{\rm eq}=\sqrt{\alpha/\beta}$. The constant $\lambda \ne 1$ in the second term of the left-hand side of Eq. (\ref{eq:toner_tu}b) marks the break-down of Galilean invariance, whereas the last term on the right-hand side is the first term in the expansion of the pressure $P=P(\rho)$. 

Following Ref. \cite{Shankar2017}, we ignore second-order differential terms, which affect the steady state configuration, but have secondary influence on the dynamics of the perturbations. In Appendix \ref{sec: TT(p)}, we consider the exact same system as that in Ref. \cite{Shankar2017}, which has modified modes. As in the previous examples we assume the active fluid to lie on a surface of revolution, whose axis is oriented in the $z-$direction and let $\phi$ and $\theta$ be the coordinate in the direction of the isometry and the perpendicular direction respectively. In Appendix \ref{sec: TT(u) general q} we consider a more general class of metrics.

Eqs. \eqref{eq:toner_tu} admit a non-trivial steady state having the same rotational symmetry of the substrate, namely:
\begin{gather*}
\rho_{0} = \rho_{\rm eq}\left[1+\frac{\lambda}{2} \left(\frac{v_{\rm eq}}{c}\right)^{2}\log\frac{r}{r_{\rm eq}}\right]\;,\\
v_{0}^{\phi} = \frac{v_{\rm eq}}{L} r^{-\frac{1}{2}}\;,\quad
v_{0}^{\theta} = 0\;,
\end{gather*}
where $r$ is defined in Eq. \eqref{eq:metric surface}. In contrast to the example illustrated in Sec. \ref{sec: shallow water}, here the steady state configuration depends on $\theta$ and does not correspond to a simple rigid body rotation of the fluid layer. Rescaling the variables:
\begin{gather*}
X = \frac{v_{\rm eq} \sqrt{r}}{c} \phi\;,\quad
Y = \frac{v_{\rm eq} \sqrt{p}}{c} \theta\;,\quad
T = \frac{2v_{\rm eq}}{L}\,t\;.\\
U = \frac{L \sqrt{r}}{2 c} \delta u^{\phi},\quad
V = \frac{L \sqrt{p}}{2 c} \delta u^{\theta}\;,\quad
H = \frac{\delta \rho}{\rho_{0}}\;,
\end{gather*}
and linearizing Eqs. \eqref{eq:toner_tu} with respect to the perturbation $\bm{\Psi}=(H,U,V)$ yields again Eq. \eqref{eq:schroedinger} with the following effective Hamiltonian:
\begin{equation}
\label{eq: H TT(u) q=0}
\bm{\mathcal{H}} = 
\begin{bmatrix} 
- \Lambda \partial_X & - \partial_X & - \partial_Y - B (r+\Delta) \\ 
-\partial_X & -\zeta & +\frac{1}{2} m \\ 
-\partial_Y &  -m &  0  
\end{bmatrix}\;,
\end{equation}
where we have set:
\begin{gather*}
m = -\lambda r^{-\frac{1}{2}} \frac{r'}{2\Delta}\;,\quad 
\zeta = L \sqrt{\alpha\beta}\;,\quad
\Lambda = \left(1-\lambda \right) \frac{v_{\rm eq}}{c}\;,\\
B = \frac{c}{2v_{\rm eq} \Delta\sqrt{p+r+2\Delta}} \left(\frac{\Delta'}{\Delta} + \frac{\rho'_0}{\rho_0} \right)\; .
\end{gather*}
The eigenvalues of $\bm{\mathcal{H}}$ for $\bm{k}=\bm{0}$ are given by:
\begin{equation} \label{eq: characteristic eq k=0}
\omega_0 = 0\;,\qquad
\omega_{\pm} = \frac{1}{2} \left( i \zeta \pm \sqrt{2 m^2 - \zeta^2} \right)\;,
\end{equation}
and the band gap closes when $m=0$ and $m=\pm \zeta/\sqrt{2}$. Eliminating $U = (k H - imV/2) / (\omega - i \zeta)$, we can again cast the equation for the remaining variables as:
\begin{equation}
\label{eq: H TT(u) q=0 2x2}
\partial_Y 
\begin{bmatrix}  
H \\ 
V 
\end{bmatrix}
=
\bm{\mathcal{H}'}
\begin{bmatrix}  
H \\ 
V 
\end{bmatrix}
\end{equation}	
where:
\begin{equation}
\bm{\mathcal{H}'}
= 
\begin{bmatrix}  
- m \frac{k}{\omega - i \zeta} & i \left( \frac{m^2}{2(\omega - i \zeta)}-\omega \right) \\ 
i \left( \Lambda k + \frac{k^2}{\omega - i \zeta} - \omega \right) & - B (r+\Delta) + \frac{k m}{2(\omega - i \zeta)} 
\end{bmatrix}	.
\end{equation}

\begin{figure*}[!t]
\centering
\includegraphics[width=\textwidth]{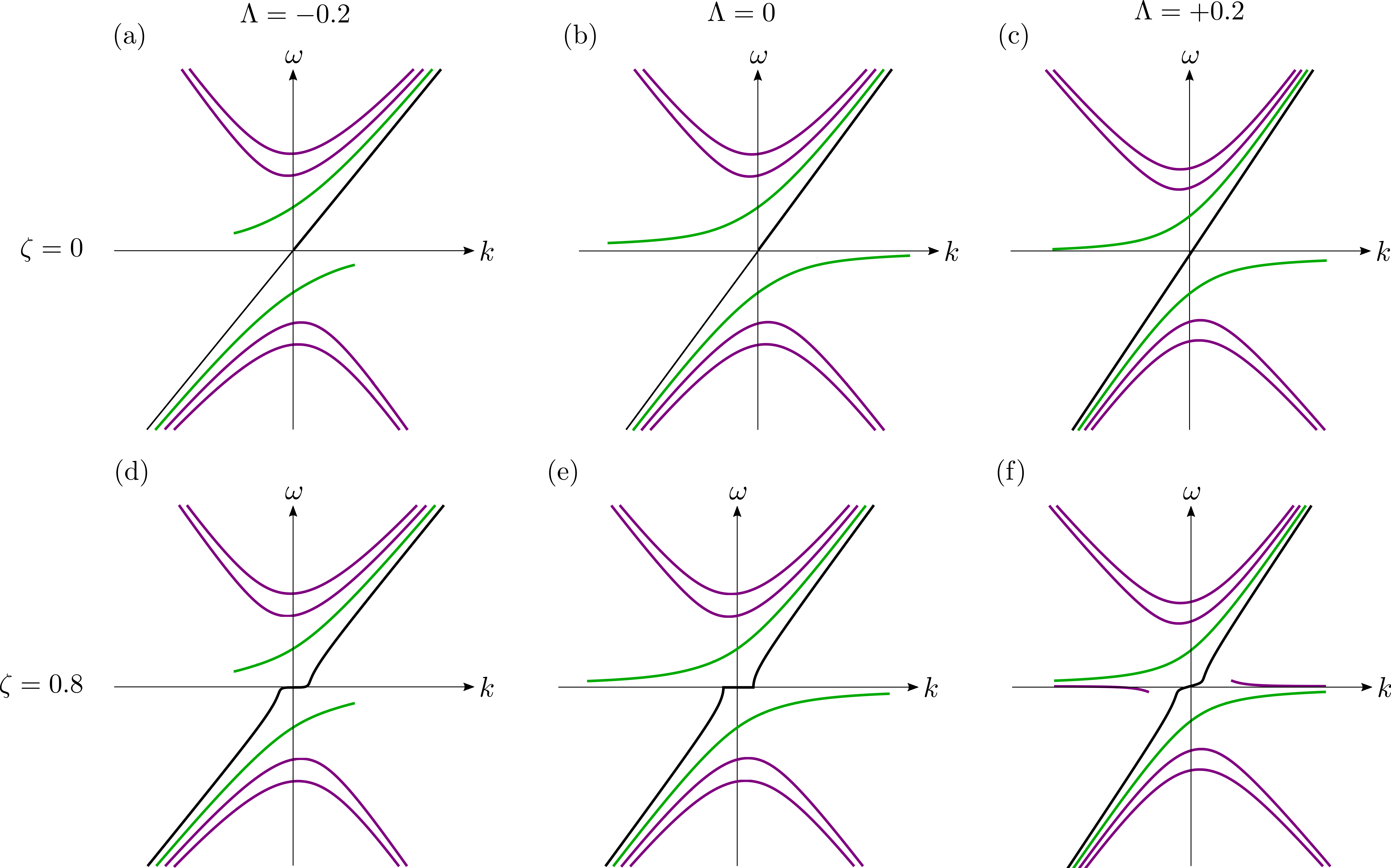}
\caption{Dispersion relations Re$(\omega)$ versus $k$ for bounded edge modes in Toner-Tu flocking for different values of $(\Lambda, \zeta)$ in the case $m'>0$. $n=0$ is shown in green, $n=1,2$ in purple and the Kelvin wave in black. Dispersion relations for $m'<0$ can be obtained by replacing $m' \to -m', \>\>\> \Lambda \to -\Lambda, \>\>\> \zeta \to \zeta, \>\>\> k \to -k, \>\>\> \omega \to \omega$.}
 \label{fig: TonerTu}
\end{figure*}

\begin{table*}[t!]
\centering
\begin{ruledtabular}  
\renewcommand{\arraystretch}{2}
\begin{tabular}{cc}
Bulk bands and topological invariants & Edge modes and topological invariants \\
$\Delta C_{\rm band} = C_{\rm band}(m>0) - C_{\rm band}(m<0)$ & $\Delta N_{\rm mode \> group} = \#({\rm right} - {\rm left})_{\rm above} - \#({\rm right} - {\rm left})_{\rm below}$ \\
\hline
\multicolumn{2}{c}{(a) Dirac fermion with position dependent mass $m(y)$} \\
\hline\\
\includegraphics[width=0.6\columnwidth]{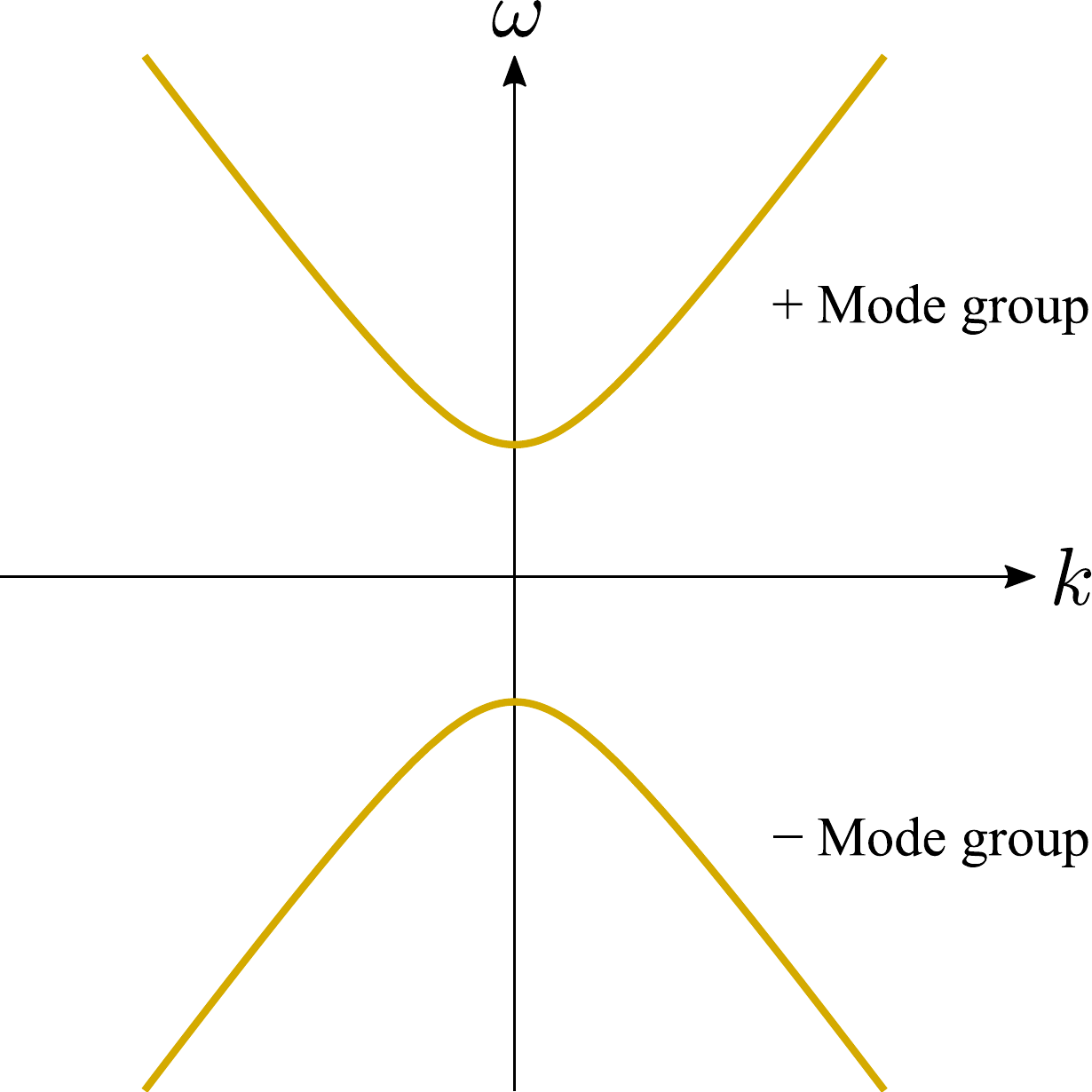} &
\includegraphics[width=0.6\columnwidth]{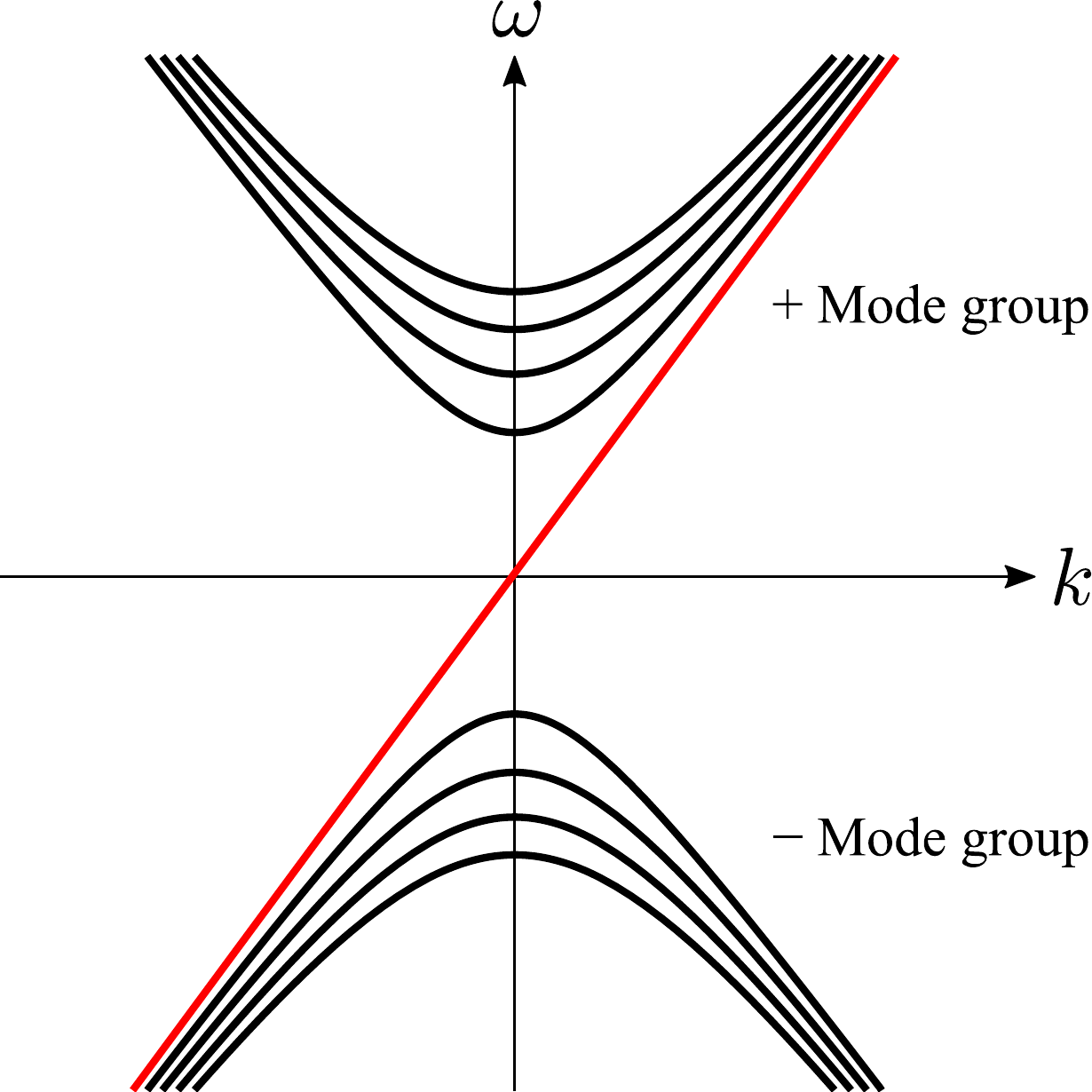} \\
$C_\pm = \mp \frac{1}{2} \left[ \sign(m) + \sign(\epsilon)
\right]$ & \\
$\Delta C_\pm = \mp 1$ & $\Delta N_{\pm} = \mp 1$ \\
\hline
\multicolumn{2}{c}{(b) Shallow water waves on a curved surface with rotating background flow} \\
\hline\\
\includegraphics[width=0.6\columnwidth]{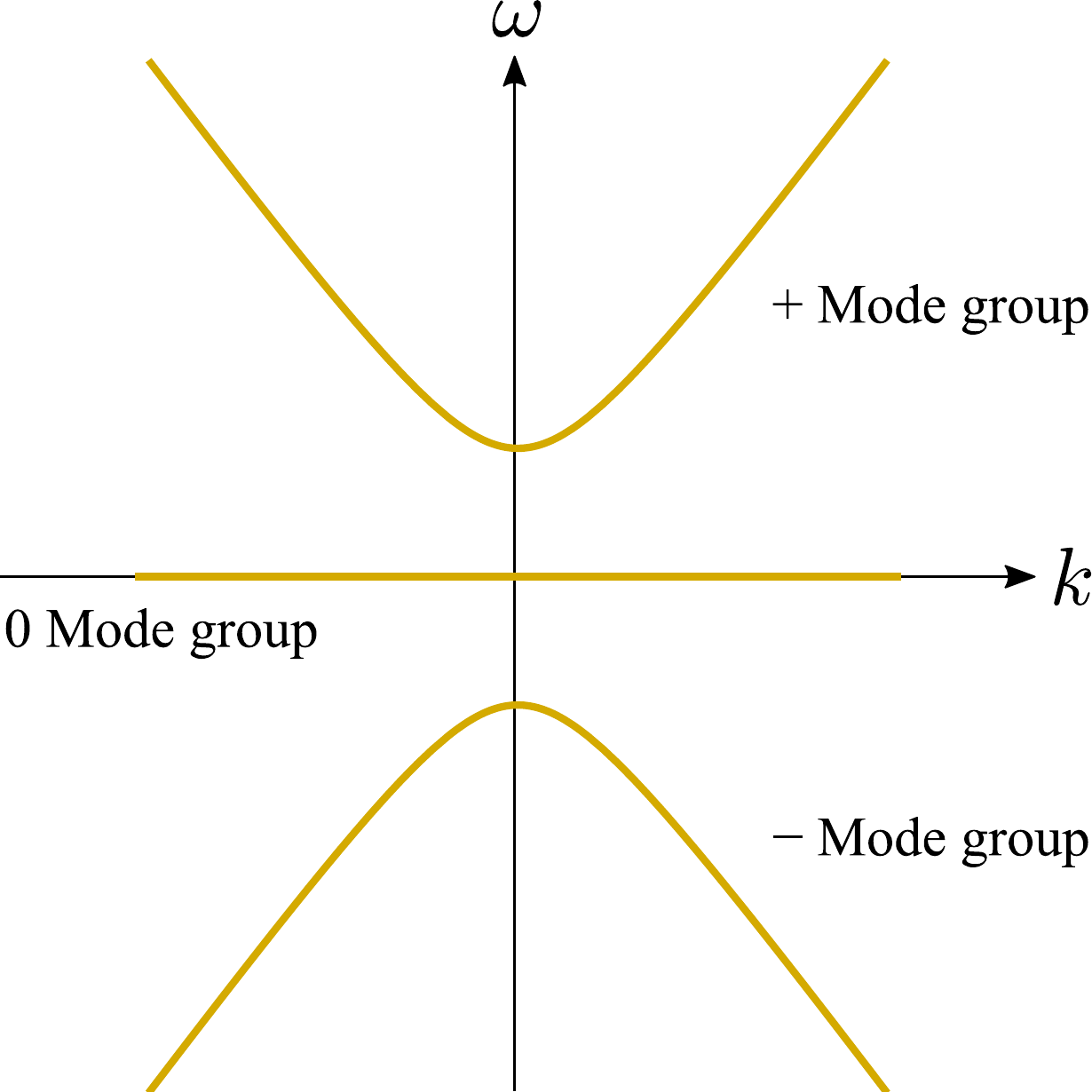} &
\includegraphics[width=0.6\columnwidth]{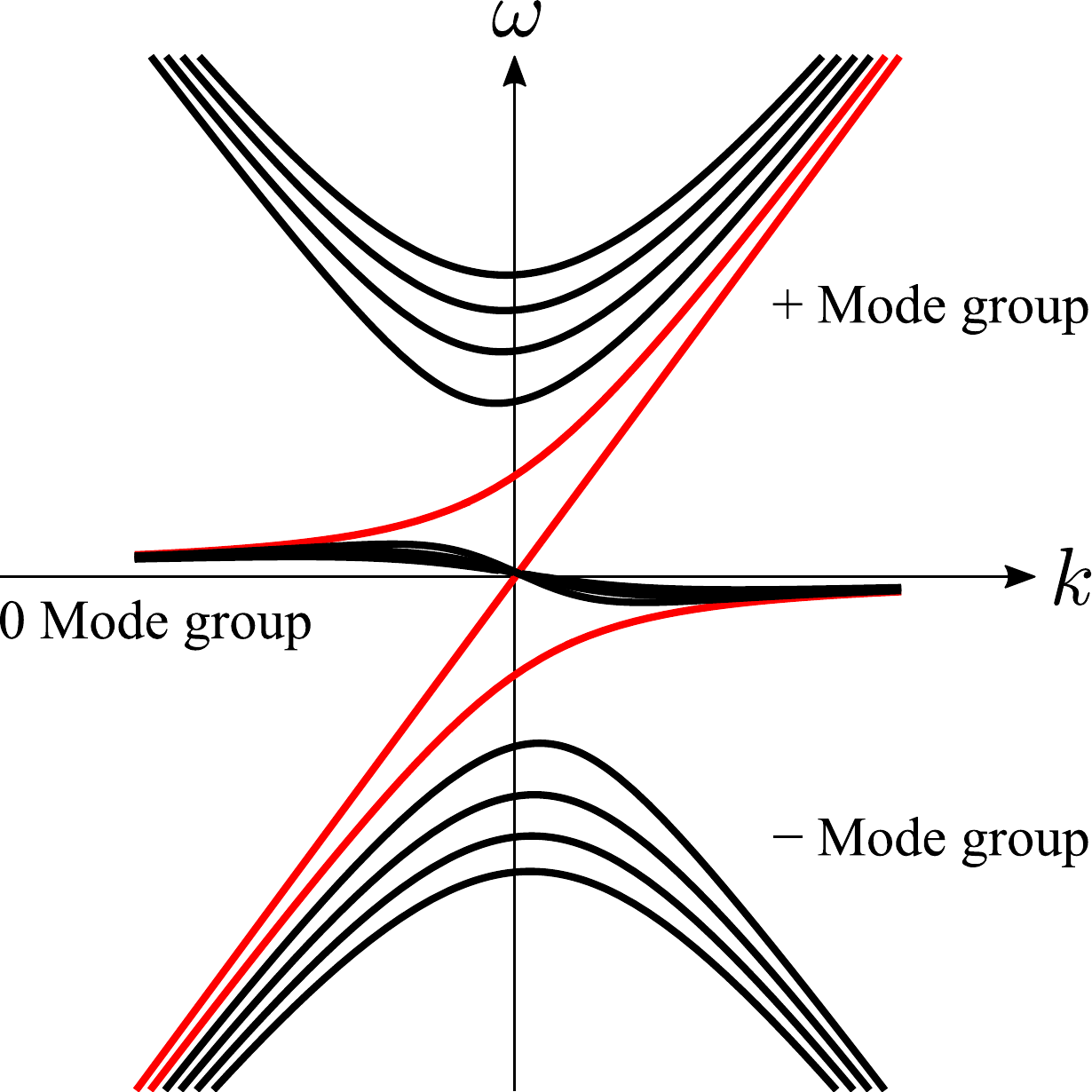} \\
$C_\pm = \mp \left[ \sign(m) + \sign(\epsilon)
\right]~, \quad C_0=0$ & \\
$\Delta C_\pm = \mp 2$~, \quad $\Delta C_0=0$ & $\Delta N_{\pm} = \mp 2$~, \quad $\Delta N_0 = 2 - 2 =0$ \\
\end{tabular}
\end{ruledtabular}
\caption{\label{fig: bulk-edge}Summary of the bulk-edge correspondence  as described in Refs. \cite{Delplace2017,Tauber_2019} for (a) Dirac fermion with position dependent mass and (b) shallow water waves on curved surface with rotating background flow. In each case, $m$ is the mass, and $\epsilon$ the regulator for high $k$ explained in the main text. Mode groups at the edge correspond to specific modes in the bulk, and for each edge mode group is associated a topological invariant defined as the net number of right-moving modes in the gap immediately above the mode group minus the net number of right-moving modes in the gap immediately below. The bulk-edge correspondence then relates this edge invariant to the change in Chern number of the relevant band across the edge. The direction of right- and left-moving modes is defined with respect to the direction of increasing $m$.}
\end{table*}

When the bottom left entry of $\bm{\mathcal{H}}'$ is zero, there can be bounded solutions of $H$ around $m=0$ with $V=0$ (for general $q$, additionally $q=0$ must be satisfied at points where $m=0$ as explained in Appendix \ref{sec: TT(u) general q}). The dispersion relation for these topological modes analogous to the Kelvin wave is then given by setting this term to zero:
\begin{equation}
\label{eq: TT(u) q=0 Kelvin wave}
\omega_{\pm} =  \frac{1}{2} \left[ \Lambda k + i \zeta \pm \sqrt{(4+\Lambda^2) k^2 - 2 i \Lambda k \zeta - \zeta^2} \right]. 
\end{equation}
For $k \ll 1$, $\omega_+ = i \zeta - ik^2 /\zeta - \Lambda(4-\Lambda^2) k^3/8 \zeta^2 +\mathcal{O}(k^4)$ and $\omega_- = \Lambda k +\mathcal{O}(k^2)$, so that when $\Lambda^2 <4$, $\omega_\mp$ is the bounded mode when sign$(m' \Lambda)=\pm 1$ respectively and the direction of propagation is given by sign$(m')$. \footnote{Note that for $m=\pm \zeta/\sqrt{2}$, if a bounded mode existed, then its dispersion relation would be given by setting ${\rm det}(\bm{\mathcal{H}'})=0$. This yields:
\begin{align*}
\frac{\zeta}{\sqrt{2}}\,k B(r+\Delta) - k^2 \omega- \left(\Lambda k - \omega \right) \left(\omega - i \frac{\zeta}{2} \right)^2 = 0\;,
\end{align*}
for which a bounded mode solution near $m(y=0)=\zeta/\sqrt{2}$ takes the form: 
\begin{align*}
\begin{bmatrix} 
H \\ 
V 
\end{bmatrix}
= 
\begin{bmatrix} 
\mu \\ 
\nu 
\end{bmatrix}
e^{-\frac{\gamma Y^{2}}{2}}\;,
\end{align*}
where $\mu$ and $\nu$ are constants. Unlike $m=0$, $\mathcal{H}'$ does not have an eigenvector which does not depend on $Y$, and so there is no bounded mode in this case.}

Eliminating $H$ in favour of $V$ in Eq. \eqref{eq: H TT(u) q=0 2x2} in the flat plane limit where $B \to 0$ and then making a change of variable $V = W \exp\{-k/[4(\omega- i \zeta)] \int {\rm d}y\,m(y)\}$, we obtain
\begin{multline}
\frac{W''}{W} 
= \frac{(\omega- \Lambda k)(\omega - i \zeta)+\frac{k^2}{8}}{2 (\omega - i \zeta)^2} m^2 \\
- \left[\omega^2 - \Lambda k \omega - \frac{k^2 \omega}{(\omega - i \zeta)} -\frac{3k}{4(\omega - i \zeta)} m' \right]\;.
\end{multline}
Taking $m=m'_0 Y$ close to $Y=0$ and rescaling $(X,Y,T, \zeta^{-1}) \to |m'_0|^{-\frac{1}{2}} (X,Y,T,\zeta^{-1})$, the dispersion relation of bounded modes is then given by
\begin{multline}
\omega^2 - \Lambda k \omega - \frac{k^2 \omega}{(\omega - i \zeta)} -\frac{3k}{4(\omega - i \zeta)} \sign (m') \\
=  \left( 2 n + 1 \right)\sqrt{ \frac{(\omega- \Lambda k)(\omega - i \zeta)+\frac{k^2}{8}}{2 (\omega - i \zeta)^2}}
\end{multline}
where $n=0,1,2 ....$, and for the mode to be bounded around $m=0$, there is a constraint on the quantity under the square root~\footnote{The solution to the equation $W'' = (A e^{i\alpha} y^2 - B e^{i\beta})W$ is $W=\exp[-\frac{1}{2} y^2 \sqrt{A} e^{i\alpha/2}] \> H_n[A^{1/4} e^{i \alpha/4}y]$, which is bounded when $\beta=\alpha/2 \in [-\pi/2,\pi/2]$ i.e. Re$[\sqrt{A e^{i \alpha}}]>0$ and $B/\sqrt{A} =2n+1$. However, we are interested in bounded modes for $V$ rather than $W$, and so the required condition is instead ${\rm Re}\{\sqrt{A e^{i \alpha}}+ \sign(m') k/4(\omega- i \zeta)\}>0$.}. The effective Hamiltonian $\bm{\mathcal{H}}$ in Eq. \eqref{eq: H TT(u) q=0 2x2} is not Hermitian and in general its eigenvalues $\omega$ are complex; the dispersion relations of Re$(\omega)$ for different values of $(\Lambda,\zeta)$ are shown in Fig. \ref{fig: TonerTu}. 
The limiting case $(\Lambda,\zeta)=(0,0)$ in Fig. \ref{fig: TonerTu}b is close to the shallow water waves problem, but with a steady-state solution for $u^\phi_0$ dependent on $\theta$, and also the net number of bounded edge modes is $+2$ for all values of $\omega$. They all have a real eigenvalue but there are no modes analogous to the low-frequency Rossby waves.
Again we see that there are two modes parallel to the flock between $\omega=0$ and the higher energy mode group as described in Sec. \ref{sec: bulk-edge} although the lower energy modes have disappeared altogether for $\Lambda \le 0, \zeta=0$ as these modes are no  longer bounded.

Since $m^{\text{flock}}=\lambda r^{-1/2} m^{\text{shallow water}}$, they have the same zeros and the same sign$(m')$, so the results in Sec. \ref{sec: shallow water general} are the same for Toner-Tu flocking, and if the hosting surface $\mathcal{M}$ is compact, then the total net number of topological modes is also given by $\chi(\mathcal{M})$.

\section{\label{sec: bulk-edge}Discussion: the bulk-edge correspondence}

In this section we follow Refs. \cite{Delplace2017,Tauber_2019,Kane2013} to provide additional details of how the bulk-edge correspondence is applied to the fluid systems in this article and discuss potential further areas of study. The bulk-edge correspondence links topological invariants in the bulk, to topological invariants on the edge, and is illustrated in Table \ref{fig: bulk-edge}. Deep in the bulk, we may take $m$ to be constant and $\bm{\mathcal{H}}$ to be translationally invariant in both $x$ and $y$. $\bm{\mathcal{H}}$ can then be expressed in terms of $k_x$ and $k_y$ and there are $n$ bands ($n=2$ for the Dirac fermion and $n=3$ for the passive and active fluids) which each have a Chern number $C$ depending on $\sign(m)$. Edges are boundaries in real space between bulk regions of different sign$(m)$, and at each edge in real space we can associate the change in bulk topological invariant, $\Delta C_i \>(i \in {1,...n})$. In the case of shallow water waves on a curved surface with rotating background flow, there are three bands in the bulk $(-,0,+)$ given by $\omega_{\pm} = \pm \sqrt{k^2+m^2}~,~\omega_0=0$ with Chern numbers $C_{\pm} =\mp \sign(m), C_0=0$. As $m$ increases from negative to positive, then $(\Delta C_-,\Delta C_0, \Delta C_+) = (+2,0,-2)$. More accurately, in the bulk at a point $Y=Y_0$ we have that $m=m(Y_0)+m'(Y_0)(Y-Y_0)+...$ and treating $m'(Y_0)$ as a correction, since $m'(Y_0)\sim\mathcal{O}(L^{-1})$, the bulk is still translationally invariant but the frequencies $\omega_{0,\pm}$ acquire corrections. Regardless, the Chern numbers associated with the bands remain the same.

The edge Hamiltonian has a slightly different structure; in the case of shallow water waves on a curved surface with rotating background flow, the $+$ and $-$ modes in the bulk become infinitely many gravity modes and the $0$ band becomes infinitely many Rossby modes satisfying the dispersion relation \eqref{eq: disp rel}. The edge topological invariant for each of these groups of modes is the net (i.e. right- minus left-traveling) number of modes in the gap above the mode group in Fourier space minus the net number of modes in the gap below it.
In our example, there are $+2$ modes between the $0$ group and $+$ group of modes, and $-2$ modes between the $-$ group and $0$ group of modes. In this case, the edge topological invariant $N_i \>(i = 0,\pm)$, being the change in net number of modes across each mode group, is $+2$ for the $-$ mode group, $0$ for the $0$ mode group, and $-2$ for the $+$ mode group \footnote{For the top and bottom mode groups, the regions respectively above and below are taken to be empty. Exceptions to this rule may be realised in other specific setups where boundary conditions play a role as in \cite{Tauber_2019}}. 
The bulk-edge correspondence then asserts that
\begin{equation}
N_i = \Delta C_i~~,
\end{equation}
for each $i = 0,\pm$.

This argument hinges on the correspondence between the mode groups in the edge Hamiltonian and the bulk modes; a natural question is, therefore, how to take a decoupling limit in which these bulk modes are removed or how to recover the bulk dynamics from the edge dynamics. Ref. \cite{Tauber_2019} found only Kelvin and Yanai waves by considering a mass proportional to the Dirac delta function $\delta(y)$ at the edge, rather than a $y-$dependent mass. However, in the systems we consider here we are not free to specify the function $m$. Instead, $m$ is fixed by the underlying geometry ($m=A_\phi$) and so a $y-$dependent mass should be considered. Additionally, the validity of hydrodynamics is limited to low frequencies and small wavenumbers $k$, thus we do not expect the dispersion relations found here to be valid for arbitrary $k$. The regime of validity can be extended by considering viscous terms in the hydrodynamic Eqs. ~\eqref{eq:generic_shallow_water}, in particular odd viscosity \cite{Tauber_2019}, to regulate the behaviour at high $k$. Introducing the regulator may shine light on how to take the appropriate decoupling limit and it will not change the topological analysis carried out here, as it is independent of the regulator. 

For non-Hermitian systems with complex eigenvalues, such as Toner-Tu flocking described in Sec. \ref{sec: TT(u)}, in order to construct an edge topological invariant, the way of partitioning modes should be extended to the topology of the complex plane. Additionally, we have not obtained the Chern numbers for the bulk bands in this context and left a more detailed analysis of the bulk-edge correspondence in active systems for future work.

\bigskip

\section{Conclusions}
\label{sec: conclusions}

In this article, we investigated the origin of topologically protected waves in classical passive and active fluids constrained on curved surfaces characterized by a ${\rm U}(1)$ isometry. Using a combination of topological band theory and real space analysis, we demonstrated that topological modes originate from a gauge-invariant component of the substrate spin connection $A_{\phi}$ that serves as the waves’ effective mass, thus influencing the waves’ band structure. In particular curves where $A_{\phi}$ changes in sign serve as internal ``edges'' for the propagation of uni-directional modes, whose propagation in the bulk is forbidden by the topological fingerprint of the band structure. For a compact surface, these occur at the maxima and minima of the distance from the surface to its symmetry axis. We have compared this with a Dirac fermion with position-dependent mass $m$, in which there is a bounded mode where $m$ changes sign and the direction of propagation is determined by $\sign(m')$. 

We then compared ``flocking'' active polar fluids to the passive shallow water systems with an external rotating drive. The systems can be made identical in the limit $\zeta \to 0, \Lambda \to 0$ if $\alpha/\beta \to \alpha r/\beta$, and we concluded that there are also a net two topological modes parallel or anti-parallel to the direction of the flock, depending on $\sign(m')$.

For compact surfaces $\mathcal{M}$, rotating shallow water waves, and Toner-Tu flocks, we then linked the number of topological modes in the band gap in momentum space to the topology of the surface $\mathcal{M}$ in real space: the net number of topological modes is given by the Euler characteristic $\chi(\mathcal{M})$ of the surface. Further analysis is required to fully understand the topology of the bands and the details of the bulk-edge correspondence in the context of active fluids. More generally, it may be possible to reformulate and generalize the index theorem presented in this article to a much larger class of problems, for example through a study of the homotopy type of the effective Hamiltonians. We hope to address this issue in the future.

\begin{acknowledgments}
We would like to thank Jasper van Wezel for useful discussions. RG is partly supported by the Delta Institute for Theoretical Physics (Delta ITP). JA is partly supported by the Netherlands Organization for Scientific Research (NWO) through the NWA Startimpuls funding scheme and by the Dutch Institute for Emergent Phenomena (DIEP) cluster at the University of Amsterdam. JdB is supported by the European Research Council under
the European Union's Seventh Framework Programme (FP7/2007-2013), ERC Grant agreement ADG 834878. LG is partially supported by Netherlands Organization for Scientific Research (NWO), through the Vidi scheme, and by the European Union through the ERC grant HexaTissues.

\end{acknowledgments}

\appendix

\begin{widetext}

\section{Derivation of shallow water equations for general metric with U(1) isometry}
\label{sec: App shallow water}

For a surface $\mathcal{M}$ with general metric
\begin{equation}
{\rm d}s^2 = L^2 \left[ p(\theta) {\rm d}\theta^2 + 2 q(\theta) {\rm d} \theta {\rm d} \phi + r(\theta) {\rm d} \phi^2 \right],
\label{eq: metric}
\end{equation}
\noindent the Christoffel symbols are given by:
\begin{eqnarray}
\Gamma^\theta_{\theta \theta} &=& \frac{r p'-2qq'}{2\Delta^2}~, \>\>\>\> \Gamma^\theta_{\theta \phi} = \frac{-q r'}{2\Delta^2}~, \>\>\>\> \Gamma^\theta_{\phi \phi} = \frac{-rr'}{2\Delta^2}~, \nonumber \\
\Gamma^\phi_{\theta \theta} &=& \frac{2pq'-qp'}{2\Delta^2}~, \>\>\>\> \Gamma^\phi_{\theta \phi} = \frac{p r'}{2\Delta^2} \>~, \>\>\>\> \Gamma^\phi_{\phi \phi} = \frac{q r'}{2\Delta^2} \>~,
\end{eqnarray}
where $'$ denotes $\partial_\theta$ and $\Delta \equiv \sqrt{p r -q^2}$. Taking a preferred orthonormal frame in which one basis vector is parallel to the direction of the isometry $\partial_\phi$ so that $e_2^\phi = 1/\sqrt{r}, e_1^\phi = -q/(\Delta \sqrt{r}), e_1^\theta = r/(\Delta \sqrt{r})$, the spin connection $A = A_\theta \> \text{d} \theta + A_\phi \> \text{d} \phi$ is then given by:
\begin{eqnarray} \label{eq:spincon}
A_\theta &=& e_1^i g_{ij} \nabla_\theta \> e_2^{j} = \frac{-q r'}{2 r \Delta} \\
A_\phi &=& e_1^i g_{ij} \nabla_\phi \> e_2^{j} = \frac{-r'}{2 \Delta}~. \nonumber
\end{eqnarray}

\noindent For a shallow water system in this general metric, the covariant equations of motion are:
\begin{eqnarray}
\partial_t + \nabla_i (h u^i) &=& 0~, \nonumber \\
\partial_t u^i + u^j \nabla_j u^i &=& -g \nabla^i h~,
\end{eqnarray}
and so a steady state solution $(h_0, u_0^i)$ in a geometry of metric (\ref{eq: metric}) satisfies:
\begin{subequations}
\begin{eqnarray}
\label{eq: st state}
\partial_\theta (h_0 u_0^\theta) + \partial_\phi (h_0 u_0^\phi) +\frac{\Delta'}{\Delta} h_0 u_0^\theta &=& 0~,  \\
\left(u_0^\theta \partial_\theta + u_0^\phi \partial_\phi \right) u_0^\phi + \frac{1}{2\Delta^2} \left[(2 p q' -q p')(u_0^\theta)^2+q r' (u_0^\phi) ^2 + 2 p r' u_0^\theta u_0^\phi \right] &=& \frac{-g}{L^2 \Delta^2} \left(p \partial_\phi - q \partial_\theta \right) h_0~, \\
\left(u_0^\theta \partial_\theta + u_0^\phi \partial_\phi \right) u_0^\theta + \frac{1}{2 \Delta^2} \left[(r p' - 2 q q')(u_0^\theta)^2 - r r' (u_0^\phi)^2 - 2 q r' u_0^\theta u_0^\phi \right] &=& \frac{-g}{L^2 \Delta^2} \left(r \partial_\theta - q \partial_\phi \right) h_0 ~.
\end{eqnarray}
\end{subequations}
When the system is rotating at angular velocity $\Omega$ in the $\phi$-direction, these equations are solved by:
\begin{equation}
u_0^\phi = \Omega, \>\>\> u_0^\theta = 0, \>\>\> h_0 = \langle h \rangle \left(1+\frac{r-\left\langle r \right\rangle}{2\alpha} \right) \equiv \langle h \rangle \> H_0~,
\end{equation}
where $\langle h \rangle$ is the average depth, $\alpha \equiv g H/(L^2 \Omega^2)$ and $\left\langle r \right\rangle$ is the average value of $r$ over the whole surface. We now linearize perturbations $ \delta h, \delta u^\phi, \delta u^\theta$ around this steady state, obtaining:
\begin{subequations}
\begin{eqnarray}
\label{eq: SWE general h}
\partial_t \> \delta h &=& -\partial_\theta (h_0 \> \delta u^\theta)-\partial_\phi (\delta h \> u_0^\phi + h_0 \> \delta u^\phi) - \frac{\Delta'}{\Delta} \> h_0 \> \delta u^\theta~, \\
\label{eq: SWE general phi}
\partial_t \> \delta u^\phi &=& \frac{-g}{\Delta^2} \> (p \partial_\phi - q \partial_\theta) \> \delta h - u_0^\phi \partial_\phi \> \delta u^\phi - (\delta u^\theta \partial_\theta + \delta u^\phi \partial_\phi) u_0^\phi  - \frac{r' u_0^\phi}{\Delta^2} \left(q \> \delta u^\phi + p  \> \delta u^\theta \right)~, \\
\label{eq: SWE general theta}
\partial_t \> \delta u^\theta &=& \frac{-g}{\Delta^2} \> (r \partial_\theta - q \partial_\phi) \> \delta h - u_0^\phi \partial_\phi \> \delta u^\theta + \frac{r' u_0^\phi}{\Delta^2} \left(r  \> \delta u^\phi + q \> \delta u^\theta \right) .
\end{eqnarray}
\end{subequations}
Rescaling as before with $H = \delta h /\langle h \rangle, t = \Omega \> T, \> (U,V)= \Omega^{-1} (\delta u^{\phi}, \> \delta u^\theta)$, and moving to a co-rotating frame $\phi \to \phi -\Omega t$ we obtain
\begin{equation}
\label{eq: SWE general}
\partial_T \begin{bmatrix} H \\ U \\ V \end{bmatrix} =
\begin{bmatrix} 0 & -H_0 \partial_\phi & -\left( H_0 \partial_\theta + \frac{r'}{2\alpha} + H_0 \frac{\Delta'}{\Delta} \right) \\ \frac{-\alpha}{\Delta^2} \left( p \partial_\phi - q \partial_\theta \right) & -\frac{r'}{\Delta^2} q & -\frac{r'}{\Delta^2} p \\ \frac{-\alpha}{\Delta^2} \left(r \partial_\theta - q \partial_\phi \right) &  \frac{r'}{\Delta^2} r &  \frac{r'}{\Delta^2} q  \end{bmatrix} \begin{bmatrix} H \\ U \\ V \end{bmatrix} ,
\end{equation}
where we have used $\partial_\theta h_0 = r' \langle h \rangle/(2\alpha)$. Now representing the metric with the matrix
\begin{equation}
\bm{\mathcal{M}} = \begin{bmatrix} r & q \\ q & p \end{bmatrix}~, \nonumber
\end{equation}
we rescale as follows:
\begin{equation}
T'=2T, \>\> \begin{bmatrix} X' \\ Y' \end{bmatrix} = 2 \alpha^{-\frac{1}{2}} \bm{\mathcal{N}}^{-1} \begin{bmatrix} \phi \\ \theta \end{bmatrix}, \>\> \begin{bmatrix} U' \\ V' \end{bmatrix} = \alpha^{-\frac{1}{2}} \bm{\mathcal{N}}^{-1} \begin{bmatrix} U \\ V \end{bmatrix},  \>\> \begin{bmatrix} \partial_{X'} \\ \partial_{Y'} \end{bmatrix} = \frac{1}{2} \sqrt{\alpha} \bm{\mathcal{N}}\begin{bmatrix} \partial_\phi \\ \partial_\theta \end{bmatrix}, \>\> H'=H~,
\end{equation}
where $\bm{\mathcal{N}}^2= \bm{\mathcal{M}}^{-1}$. The two bottom rows  of Eq. (\ref{eq: SWE general}) become:
\begin{equation} 
\partial_{T'} \begin{bmatrix} U' \\ V' \end{bmatrix} = \begin{bmatrix}
-\frac{\sqrt{\alpha}}{2} \bm{\mathcal{N}} \begin{bmatrix} \partial_\phi \\ \partial_\theta \end{bmatrix} & \frac{-r'}{2 \Delta^2} \bm{\mathcal{N}}^{-1} \begin{bmatrix}q & p \\ -r & -q \end{bmatrix} \bm{\mathcal{N}}
\end{bmatrix} \begin{bmatrix} H' \\ U' \\ V' 
\end{bmatrix} = \begin{bmatrix} -\partial_{X'} & 0 & \frac{-r'}{2 \Delta} \\ -\partial_{Y'} & \frac{r'}{2 \Delta} & 0 \end{bmatrix} \begin{bmatrix} H' \\ U' \\ V' \end{bmatrix} ,
\end{equation}
where we have used
\begin{equation}
\bm{\mathcal{N}}=\frac{1}{\Delta \sqrt{p+r+2 \Delta}} \begin{bmatrix} p+ \Delta & -q \\ -q & r+\Delta
\end{bmatrix}, \>\> \bm{\mathcal{N}}^{-1} =\frac{1}{\sqrt{p+r+2 \Delta}} \begin{bmatrix} r+ \Delta & q \\ q & p+\Delta
\end{bmatrix} . \nonumber
\end{equation}
For the top line of Eq. (\ref{eq: SWE general}),
\begin{equation}
\partial_{T'} \> H' = \left[ - \frac{H_0}{2} \begin{bmatrix}\partial_\phi \\ \partial_\theta \end{bmatrix} - \frac{1}{2}\begin{bmatrix} 0 \\ H_0 \frac{\Delta'}{\Delta} + \frac{r'}{2\alpha} \end{bmatrix} \> \right]^{\intercal} \sqrt{\alpha} \bm{\mathcal{N}} \begin{bmatrix} U' \\ V' \end{bmatrix} = \left[ - H_0 \begin{bmatrix}\partial_X \\ \partial_Y \end{bmatrix} - \frac{\sqrt{\alpha} \left(H_0 \frac{\Delta'}{\Delta}+\frac{r'}{2\alpha} \right)}{2 \Delta \sqrt{p+r+2\Delta}} \begin{bmatrix} -q \\ r+\Delta \end{bmatrix} \> \right]^{\intercal} \begin{bmatrix} U' \\ V' \end{bmatrix} ,
\end{equation}
and provided we restrict our attention to regions of the surface where $H_0 \approx 1$ (for small fluctuations in $r$), rewriting $m=A_\phi=-r'/(2\Delta)$ from Eq. \eqref{eq:spincon} we obtain
\begin{eqnarray}
\partial_{T'} \begin{bmatrix} H' \\ U' \\V' \end{bmatrix} &=& \begin{bmatrix} 
0 & - \partial_X + B q & - \partial_Y - B \left(r+\Delta \right) \\
-\partial_X & 0 & m \\ -\partial_Y & -m & 0
\end{bmatrix} 
\begin{bmatrix}
H' \\ U' \\ V'
\end{bmatrix}~, \\
\text{with} \>\>\>\> B &\equiv & \frac{\sqrt{\alpha}}{2\Delta\sqrt{p+r+2\Delta}} \left(\frac{r'}{2\alpha}+ \frac{\Delta'}{\Delta} \right)~. \nonumber
\end{eqnarray}
When $q=0$ this agrees with Eq. \eqref{eq:h_u1}. Noting that $B\sim \mathcal{O}\left(L^{-1}\right)$, the $B$ terms above can be neglected when looking at small enough regions. Thus, the analysis of Sec.~\ref{sec: shallow water general} carries through even when $q\ne0$.

\section{Toner-Tu flocking in the case of general $q$}
\label{sec: TT(u) general q}

For general metric $\text{d} s^2 = L^2 \left[ p(\theta) \text{d} \theta^2 + 2 q(\theta) \text{d} \theta \text{d} \phi + r(\theta) \text{d} \phi^2 \right]$, we work with the covariant form of eq. \eqref{eq:toner_tu}:
\begin{eqnarray}
\partial_t \rho + \nabla_i ( \rho u^i) &=& 0 \nonumber~, \\
\partial_t u^i + \lambda u^j \nabla_j u^i &=& \left(\alpha - \beta g_{j k} u^j u^k \right) u^i - \frac{c^2}{\rho_{\text{eq}}} \nabla^i \rho ~ . \>\>\>\>\>\>
\end{eqnarray}
\noindent The steady state is the same as for the case $q=0$ worked in the main text. Linearizing with perturbations $\delta \rho, \delta u^\phi, \delta u^\theta$ around the steady state in which $u^{\phi}_0 \equiv \Omega(\theta)$ and moving to a co-rotating frame $\phi \to \phi -\lambda \Omega t$ where again we confine our attention to a region in which we can take this to be a rigid body rotation, gives
\begin{equation}
\partial_t \begin{bmatrix} \delta \rho \\ \delta u^{\phi} \\ \delta u^{\theta} \end{bmatrix} = \begin{bmatrix} \Omega (\lambda-1) \partial_\phi & -\rho_0 \partial_\phi & - \rho_0 \left(\partial_\theta + \frac{\rho'_0}{\rho_0}+\frac{\Delta'}{\Delta} \right) \\ \frac{-c^2}{L^2 \Delta^2 \rho_{\text{eq}}} \left( p \partial_\phi - q \partial_\theta \right) & - \lambda \frac{r' \Omega}{\Delta^2} q -2 \alpha & - \lambda \frac{r' \Omega}{\Delta^2} \left(p - \frac{\Delta^2}{2 r} \right) -2 \alpha  \frac{q}{r} \\ \frac{-c^2}{L^2 \Delta^2 \rho_{\text{eq}}} \left(r \partial_\theta - q \partial_\phi \right) &  +\lambda \frac{r' \Omega}{\Delta^2} r &  +\lambda \frac{r' \Omega}{\Delta^2} q  \end{bmatrix} 
\begin{bmatrix} \delta \rho \\ \delta u^{\phi} \\ \delta u^{\theta} \end{bmatrix}.
\end{equation}
Combining rescalings as before with
\begin{equation}
T=\frac{2 c}{L} t, \>\>\> \begin{bmatrix} X \\ Y \end{bmatrix} = \frac{v_{\rm eq}}{c} \bm{\mathcal{N}}^{-1} \begin{bmatrix} \phi \\ \theta \end{bmatrix},  \>\>\> \begin{bmatrix} U \\ V \end{bmatrix} = \frac{L}{2 c} \bm{\mathcal{N}}^{-1} \begin{bmatrix} \delta u^{\phi} \\ \delta u^{\theta} \end{bmatrix}, \>\>\> H=\frac{\delta \rho}{\rho_0}
\end{equation}
then leads to
\begin{eqnarray}
\bm{\mathcal{H}} &=& \begin{bmatrix} - \Lambda\frac{(r+\Delta) \partial_X + q \partial_Y}{\sqrt{r(p+r+2\Delta})} & \frac{\rho_0}{\rho_{\text{eq}}} \left(- \partial_X + B q \right) & \frac{\rho_0}{\rho_{\text{eq}}} \left[- \partial_Y - B (r+\Delta) \right] \\ -\partial_X  & 0 & +m \\ -\partial_Y &  -m &  0  \end{bmatrix} \nonumber \\
&& \>\>\>  - \frac{m}{2 r (p+r+2\Delta)} \begin{bmatrix} 0 & 0 & 0 \\ 0 & - q (r+\Delta) & (r+\Delta)^2  \\ 0 & -q^2 &  q (r+\Delta)  \end{bmatrix}  - \frac{\zeta}{r (p+r+2\Delta)} \begin{bmatrix} 0 & 0 & 0 \\ 0 & (r+\Delta)^2 & q (r+\Delta) \\ 0 &  q (r+\Delta) & q^2  \end{bmatrix}~,
\end{eqnarray}
where now we have set
\begin{equation}
\>\>\>\ m = -\lambda r^{-\frac{1}{2}} \frac{r'}{2\Delta}, \>\>\> \zeta = L \sqrt{\alpha \beta}, \>\>\> B = \frac{c}{2 v_{\rm eq} \Delta \sqrt{p+r+2\Delta}} \left(\frac{\Delta'}{\Delta} + \frac{\rho'_0}{\rho_0} \right), \>\>\> \Lambda = \left(1-\lambda \right) \frac{v_{\rm eq}}{c}~. \nonumber
\end{equation}
\noindent Eliminating $U$ and taking into account the possible dependence of $\omega$ on $Y$ we obtain

\begin{eqnarray}
\partial_Y H - i (\partial_Y \omega) T H &=& H \left\lbrace k \frac{m \left[ -1 + \frac{q^2}{2r(p+r+2\Delta)} \right]-\frac{q}{r} \frac{r+\Delta}{p+r+2\Delta} \zeta}{\omega - i \zeta + \frac{i q \left[m(r+\Delta) + 2 q \zeta \right]}{2 r (p+r+2\Delta)}} \right\rbrace + i V \left\lbrace \frac{\left[\frac{m^2}{2(\omega - i \zeta)} - \omega \right]}{1+ i \frac{q}{r} \frac{m(r+\Delta) + 2 q \zeta}{2 r (p+r+\Delta) (\omega - i \zeta)}} \right\rbrace \nonumber \\
\partial_Y V - i (\partial_Y \omega) T V &=& i H \left\lbrace \frac{\Lambda k  \left[r+\Delta + q \frac{-m q^2 + 2 m r (p+r+2\Delta) + 2 q(r+\Delta) \zeta}{-mq(r+\Delta) - 2 q^2 \zeta + 2 i r (p+r+\Delta) (\omega - i \zeta)} \right]}{\sqrt{r(p+r+2\Delta)}} + \frac{\left(k^2 - i q \frac{B k}{2} \right)}{\omega - i \zeta + i q \frac{m(r+\Delta) +2 q \zeta}{2 r (p+r+2 \Delta)}} - \omega \right\rbrace \nonumber \\
&&+V \left\lbrace -B (r+\Delta) +k \frac{m \left[ \frac{1}{2} + \frac{q^2}{2r(p+r+2\Delta)} \right]-\frac{q}{r} \frac{r+\Delta}{p+r+2\Delta} \zeta}{\omega - i \zeta + \frac{i q \left[m(r+\Delta) + 2 q \zeta \right]}{2 r (p+r+2\Delta)}} - \frac{i \Lambda q}{r} \frac{\sqrt{r(p+r+2\Delta)} \left[\frac{m^2}{2(\omega - i\zeta)} - \omega \right]}{(p+r+2\Delta) + i \frac{q}{r} \frac{m(r+\Delta) + 2 q \zeta}{2(\omega - i \zeta)}} \right\rbrace . \nonumber \\
\label{eq: monster}
\end{eqnarray}
As a check, note that Eq. (\ref{eq: monster}) reduces to Eq. (\ref{eq: H TT(u) q=0 2x2}) when $q \equiv 0$. Now the Kelvin wave edge mode is shifted to points at which
\begin{equation}
m \left[ -1 + \frac{q^2}{2r(p+r+2\Delta)}\right]-\frac{q}{r} \frac{r+\Delta}{p+r+2\Delta} \zeta = 0~,
\end{equation}
instead of $m=0$. The dispersion relation
\begin{equation}
\frac{\Lambda k  \left[r+\Delta + q \frac{-m q^2 + 2 m r (p+r+2\Delta) + 2 q(r+\Delta) \zeta}{-mq(r+\Delta) - 2 q^2 \zeta + 2 i r (p+r+\Delta) (\omega - i \zeta)} \right]}{\sqrt{r(p+r+2\Delta)}} + \frac{\left(k^2 - i q \frac{Bk}{2} \right)}{\omega - i \zeta + i q \frac{m(r+\Delta) +2 q \zeta}{2 r (p+r+2 \Delta)}} - \omega = 0~,
\end{equation}
then depends on $Y$ and the direction of the bounded mode is determined by the sign of
\begin{equation}
\partial_Y \left\lbrace \frac{m \left[ -1 + \frac{q^2}{2r(p+r+2\Delta)} \right]-\frac{q}{r} \frac{r+\Delta}{p+r+2\Delta} \zeta}{\omega - i \zeta + \frac{i q \left[m(r+\Delta) + 2 q \zeta \right]}{2 r (p+r+2\Delta)}} \right\rbrace~, \nonumber
\end{equation}
instead of simply $\partial_Y m$. In the case of the helicoidal sphere shown in Fig. \ref{fig: shapes_modes}(d), $m=0$ the helix $\theta=\pi/2$, but here $q \ne 0$. In this geometry therefore, the activity parameter $\zeta>0$ shifts the bounded topological edge modes upwards compared with the $\zeta=0$ case.

\section{Alternative model for active polar fluids}
\label{sec: TT(p)}

In Ref. \cite{Shankar2017}, the flocking system is described by an alternative model of active polar fluids, described by the following set of hydrodynamic equations for the density $\rho$ and momentum density $\bm{p}=\rho\bm{v}$:
\begin{subequations}\label{eq: T-Tu(p)}
\begin{gather}
\partial_t \rho + \nabla \cdot \bm{p}  = 0\;, \\
\partial_t \bm{p} + \lambda_1 \bm{p} \cdot \nabla \bm{p} = \left[ \alpha_1 (\rho - \rho_c) - \beta_1 |\bm{p}|^2 \right] \bm{p} - c^2 \nabla \rho\;.
\end{gather}
\end{subequations}
While the constant $\lambda$ in the main text is dimensionless, $\lambda_1$ has dimension $\rho^{-1}$ and this, along with the dependence of the drive on $\rho$, affects the steady state and Fourier modes of perturbations in the system. With rescaled variables
\begin{equation}
T =  \frac{2 v_1 t}{L \sqrt{\lambda_1 (\rho_{\text{eq}} -\rho_c ) r_{\text{eq}}}} , 
\> \begin{bmatrix} X \\ Y \end{bmatrix} = \frac{2 v_1 \sqrt{\lambda_1 (\rho_{\text{eq}} -\rho_c)r_{\text{eq}}}}{c} \bm{\mathcal{N}}^{-1} \begin{bmatrix} \phi \\ \theta \end{bmatrix}, \> \begin{bmatrix} U \\ V \end{bmatrix} = \frac{1}{c (\rho_{\text{eq}} -\rho_c)} \bm{\mathcal{N}}^{-1} \begin{bmatrix} \delta p^\phi \\ \delta p^\theta \end{bmatrix}, H = \frac{\delta \rho}{\rho_{\text{eq}} -\rho_c}~,
\end{equation}
the effective Hamiltonian for a diagonal metric ($q=0$) is now:
\begin{eqnarray}
\label{eq: Shankar}
\bm{\mathcal{H}}_1 &=& \begin{bmatrix} \Lambda_1 \partial_X & - \partial_X & - \partial_Y - B_1 (r+\Delta) \\ -\partial_X + D_1 \zeta_1 & -\zeta_1 & +m_1 \frac{1+\eta_1}{2}\\ -\partial_Y &  -m_1 &  0  \end{bmatrix}~,
\end{eqnarray}
where
\begin{gather*}
\eta_1 = \frac{\lambda_1 \alpha_1}{\beta_1 v_1}\;,\quad 
 m_1 = - \frac{\lambda_1 ( \rho_{\text{eq}} -\rho_c ) r'}{2\Delta} \left( \frac{r}{r_{\text{eq}}} \right)^{\frac{\eta_1}{2} -1}\;,\quad
\zeta_1 = L \sqrt{\alpha_1 \beta_1 r_{\text{eq}} (\rho_{\text{eq}} -\rho_c)^3} \left( \frac{r}{r_{\text{eq}}} \right)^{\frac{\eta_1}{2}}\;, \\
B_1 = \frac{c}{v_1} \sqrt{\frac{\lambda_1 (\rho_{\text{eq}}-\rho_c) r_{\text{eq}}}{(p+r+2\Delta)}}\frac{\Delta'}{2\Delta^2}\; , \quad \Lambda_1 = \frac{v_1}{c} \sqrt{\lambda_1 (\rho_{\text{eq}} - \rho_c)}\left(\frac{r}{r_{\text{eq}}}\right)^{\frac{\eta_1}{4}}\; , \quad D_1 = \frac{v_1}{2 c \sqrt{\lambda_1(\rho_{\text{eq}} - \rho_c) r_{\text{eq}}}} \left(\frac{r}{r_{\text{eq}}}\right)^{-\frac{\eta_1}{4}} ~.
\end{gather*}
Comparing Eq. \eqref{eq: Shankar} to Eq. \eqref{eq: H TT(u) q=0} for the original Toner-Tu equations in the main text, $\Lambda_1 \sim \lambda_1$ whereas $\Lambda \sim (1-\lambda)$, which is a result of modes being described in terms of $\delta \bm{p}$ instead of $\delta \bm{v}$. Also in the $\phi$-equation there is a a new factor $D_1 \zeta_1$ coupling to the density perturbation, which comes from the drive's dependence on $\rho$, and an additional factor of $m_1 \eta_1/2$ coupling to $\delta p^{\theta}$, which follows from the steady state's modified dependence on $\theta$.
Eliminating $U$ from Eq. (\ref{eq: Shankar}) gives
\begin{equation}
\bm{\mathcal{H}}'_1 = \begin{bmatrix}  - m_1 \frac{k- i D_1 \zeta_1}{\omega - i \zeta_1} & i \left[ \frac{m^2 (1+\eta_1)}{2(\omega - i \zeta_1)}-\omega \right] \\ i \left[- \Lambda_1 k + \frac{k(k-i D_1 \zeta_1)}{\omega - i \zeta_1} - \omega \right] & - B_1 (r+\Delta) + \frac{k m (1+\eta_1)}{2(\omega - i \zeta_1)} \end{bmatrix} ~,
\end{equation}
which has a modified dispersion relation
\begin{equation}
\omega_{\pm} =  -\frac{1}{2} \left\lbrace \Lambda_1 k - i \zeta_1 \pm \sqrt{(-\Lambda_1 k + i \zeta_1 )^2 + 4 \left[k^2+i \zeta_1 k (\Lambda_1-D_1) \right]} \right\rbrace ~.
\end{equation}
Now for $k \ll 1$ close to zero, 
$\omega_- \approx i \zeta_1 - D_1 k$, $ \omega_+ \approx (D_1-\Lambda_1) k$, and the mode for $H$ with $V=0$ is bounded when 
\begin{equation}
- m'_1 \frac{k- i D_1 \zeta_1}{\omega - i \zeta_1} < 0~.
\end{equation}
This is satisfied for $\omega_\pm$ respectively when sign$(m'_1)=\pm 1$. When $m'_1>0$, the Kelvin wave travels in the sign$(D_1-\Lambda_1)$ direction, so is parallel to the flock when $D_1>\Lambda_1$ and counter to the flock when $D_1<\Lambda_1$. When $m'_1<0$ however, the Kelvin wave always travels counter to the flock, i.e. opposite to that given in the description of the catenoid in Ref. \cite{Shankar2017}. 

\end{widetext}

\bibliography{bibliography}

\end{document}